\documentclass[12pt, twocolumn]{aastex61}
\usepackage{natbib}
\usepackage{graphicx}

\newcommand{\chisq}{\chi_{\rm red}^2}

\begin{document}
	
	\title{Thermal Modeling of Comet-Like Objects from AKARI Observation}
	
	\author{Yoonsoo P. BACH}
	\affil{Department of Physics and Astronomy, Seoul National University,
		Gwanak, Seoul 151--742, South Korea}
	\email{ysbachpark@astro.snu.ac.kr}
	
	\author{Masateru ISHIGURO}
	\affil{Department of Physics and Astronomy, Seoul National University,
		Gwanak, Seoul 151--742, South Korea}
	\email{ishiguro@astro.snu.ac.kr}
	
	\author{Fumihiko USUI}
	\affil{Center for Planetary Science, Graduate School of Science, Kobe University,
		7-1-48, Minatojima-Minamimachi, Chuo-Ku, Kobe 650-0047, Japa}
	\email{ishiguro@astro.snu.ac.kr}

	\begin{abstract}
		We investigated the physical properties of the comet-like objects 107P/(4015) Wilson--Harrington (4015WH) and P/2006 HR30 (Siding Spring; HR30) by applying a simple thermophysical model (TPM) to the near-infrared spectroscopy and broadband observation data obtained by AKARI satellite of JAXA when they showed no detectable comet-like activity. We selected these two targets since the tendency of thermal inertia to decrease with the size of an asteroid, which has been demonstrated in recent studies, has not been confirmed for comet-like objects. It was found that 4015WH, which was originally discovered as a comet but has not shown comet-like activity since its discovery, has effective size $ D= $ 3.74--4.39 km and geometric albedo $ p_V \approx $ 0.040--0.055 with thermal inertia $ \Gamma = $ 100--250 J m$ ^{-2} $ K$ ^{-1} $ s$ ^{-1/2}$. The corresponding grain size is estimated to 1--3 mm. We also found that HR30, which was observed as a bare cometary nucleus at the time of our observation, have $ D= $ 23.9--27.1 km and $ p_V= $0.035--0.045 with $ \Gamma= $ 250--1,000 J m$ ^{-2} $ K$ ^{-1} $ s$ ^{-1/2}$. We conjecture the pole latitude $ - 20^{\circ} \lesssim \beta_s \lesssim +60^{\circ}$. The results for both targets are consistent with previous studies. Based on the results, we propose that comet-like objects are not clearly distinguishable from asteroidal counterpart on the $ D $--$ \Gamma $ plane. 
	\end{abstract}
	
	
\section{Introduction}\label{sec:introduction}

Any material with non-zero absolute temperature emits \emph{black body radiation}, and its irradiation is described by Plank's law in monochromatic intensity units (W/m$ ^2 $/$ \mu $m/sr; \citealt{Planck1914}):
\begin{equation}\label{eq: Planck law}
B_\lambda(\lambda,T) = \frac{2 h c^2}{\lambda^5} \frac{1}{e^{hc/\lambda k T} - 1} ~.
\end{equation}
The notations for the physical constants and variables used in this paper are summarized in Table \ref{tab: notation}. The black body radiation flux dominates the reflected sunlight component at longer wavelength region, which is called the thermal region (generally $ \lambda \gtrsim 3-4 ~\mu$m for inner Solar System bodies).


Including the estimation of the size and Bond albedo of asteroid (4) Vesta \citep{Allen1970}, researchers started investigating the thermal flux in the mid 20th century. The pioneering model is referred to as the standard thermal model (STM; \citealt{Morrison1979, Lebofsky1986a, Lebofsky1989}). This STM succeeded in determining the size and albedo sets of many main-belt asteroids, and have been widely used, especially when dealing with large datasets from infrared surveys, such as IRAS \citep{Tedesco2002}, and AKARI \citep{Usui2011a}. An updated version of STM, called the near-Earth asteroids thermal model (NEATM; \citealt{Harris1998}), has been used for later surveys, e.g., WISE \citep{Mainzer2011c}. However, both STM and NEATM assume instantaneous thermal equilibrium with insolation, i.e., zero thermal inertia, so night-emission is completely ignored. Even its variants can only deal with non-rotating, fast-rotating, or infinitely high thermal inertia cases. 

Thermal inertia ($ \Gamma$)\footnote{$ \Gamma$ has the SI unit J m$ ^{-2} $ K$ ^{-1} $ s$ ^{-1/2} $. We will call this ``SI'' for shorthand. Depending on the authors, ``MKS'' or ``tiu'' (first proposed by \citealt{Putzig2006}) are occasionally used as alternates.} is a quantity that measures the thermal conduction efficiency, which is defined as
\begin{equation} \label{eq: TI def}
\Gamma \equiv \sqrt{\kappa \rho c_s} ~.
\end{equation}
Although some studies, such as \cite{Dickel1979}, modeled thermal emissions for some finite $ \Gamma$ values, \cite{Spencer1989} provided one of the first successes in developing a useful yet simple thermal model formalism for using $ \Gamma$ as a free parameter. Now, any variant of this kind is called a thermophysical model (TPM).

The parameter $ \Gamma$ can be determined from applying TPM, and it is a key parameter not only to model the dynamic evolution of small bodies in the long term but to obtain clues about the physical properties of its surface. From a dynamic viewpoint, $ \Gamma$ controls the Yarkovsky and Yarkovsky--O'Keefe--Radzievskii--Paddack (YORP) effects, which change the orbital elements and the pole orientation in the long term, respectively \citep{Vokrouhlicky2015}. It also provides hints about the boulder size on the surface (\citealt{Gundlach2013}; \citealt{Delbo2015a} and references therein) and the regolith formation of small asteroids \citep{Delbo2014}. This information also has tremendous implications for planning space missions. 

Since $ \Gamma$ is a monotonically increasing function of thermal conductivity, a lower $ \Gamma$ value may indicate ineffective thermal conduction from the solar-heated top layer to deeper regions, so the environment is favorable for ice to survive for a longer time. As a consequence, it has long been suggested that cometary bodies should have very low $ \Gamma$ compared to that of asteroidal bodies, and the preliminary results from the Deep Impact mission appeared to strongly support this idea: \cite{AHearn2005} suggested the $ \Gamma$ of comet 9P/Tempel 1 nucleus to be less than 100 SI, while similar-sized asteroidal counterparts generally have $ \Gamma$'s greater than 100 SI \citep{Delbo2015a}. Theoretical studies also showed that the survival rate of water ice depends critically on the average temperature of a small body. The temperature is a strong function of $ \Gamma $ value, as well as orbit, grain size and porosity, and spin orientation \citep{Schorghofer2008}.

Later, however, detailed thermal modeling on the comet 9P/Tempel 1 was performed by \cite{Davidsson2013}, and the best-fitting $ \Gamma$ was found to be as high as 250 SI, depending on the region, although \cite{Groussin2013} calculated it to be less than 45 SI. \cite{Groussin2013} also calculated the $ \Gamma$ of comet 103P/Hartley 2 nucleus to be $ \lesssim $ 250 SI based on EPOXI mission observations, which still leaves the possibility of a higher $ \Gamma$ (hundreds of SI) for comet nuclei. 

It is thus of great importance to have observational data of the bare nucleus of cometary bodies, as well as asteroids, to investigate the possible difference in $ \Gamma $ value between these populations. Except for mission targets, however, there is no available open spectroscopic observation data of comets. This is mainly because it is extremely difficult to observe comets to obtain useful data for thermal modeling since cometary activity can easily veil the nucleus itself. Few such precious observations were made successfully by AKARI satellite: the comet P/2006 HR30 (Siding Spring) and the comet-like asteroid 107P/(4015) Wilson--Harrington (HR30 and 4015WH hereafter) were observed without visible comet-like activity. Although HR30 showed clear cometary activity near its apparition in 2006 \citep{Hicks2007}, we confirmed that it was inactive at the time of the AKARI observations in January of 2007 based on visual inspection and comparison with the stellar point spread function (See Section \ref{sec:Observation}). 

4015WH exhibited one-time cometary activity but never again \citep{Fernandez1997,Ishiguro2011}. It has a low probability of Jupiter-family comet's origin (i.e. $ \sim 4 $ \%, \citealt{Bottke2002}). A detailed investigation using Spitzer Space Telescope has also been conducted applying NEATM \citep{Licandro2009}. They derived $ D = 3.46 \pm 0.32 $ km, $ p_V = 0.059 \pm 0.011 $, and $ \eta = 1.39 \pm 0.26 $). The corresponding lower limit for $ \Gamma $ was given as 60 SI. Since 4015WH is one of the small bodies in the near-Earth region that are accessible by existing spacecrafts, it has been considered as a sample return mission target \citep{Kawaguchi2002,Barucci2009,Yoshikawa2008}. Obtaining reliable knowledge on near-Earth objects is also directly related to human beings. We must know the physical properties, including the size, surface material strength, and composition of objects, especially for those that approach close to Earth, for the realization of future planetary defense technology, such as asteroid deflection or disruption \citep{Wie2013,Kaplinger2013}. 

Meanwhile, the target HR30 is a comet that did not show any visible cometary activity during spectroscopic observations, which is a very rare opportunity to study physics of a cometary nucleus. Moreover, it is one of the largest known potentially-dormant comets \citep{Kim2014}.

In Section \ref{sec:Observation}, we describe the AKARI observation data and reduction process. Then we describe the model and its implementation in Section \ref{sec:Methodology}. The results using TPM are summarized in Section \ref{sec:Results}, and the corresponding discussions are given in Section \ref{sec:Discussion}. 

For brevity, we summarize the notations for the constants and variables in Table \ref{tab: notation}. 

\begin{deluxetable}{lll} 
	\linespread{0.8}
	\tablewidth{0pt} 
	\tablecaption{Symbols used in this paper. \label{tab: notation}}
	\tabletypesize{\scriptsize}
	\tablehead{\colhead{\textbf{Symbol}}  & \colhead{\textbf{Description}}  & \colhead{\textbf{Value and Unit}} }
	\startdata
	$ c $              & Speed of light                        & $ 2.998 \times 10^8 $ m/s         \\
	$ h $              & Planck's constant                     & $ 6.626 \times 10^{-34} $ J s         \\
	$ k $              & Boltzmann constant                    & $ 1.381 \times 10^{-23} $ J/K         \\
	$ L_\sun $         & Solar luminosity                      & $ 3.828 \times 10^{26} $ W           \\
	$ V_{\sun} $       & Visual magnitude of the Sun           & $ -26.762 $ (mag)       \\
	$ \sigma_{\rm SB} $& Stefan--Boltzmann constant            & $ 5.670 \times 10^{-8} $ W m$ ^{-2} $ K$ ^{-4} $ \\
	\hline
	$ A_B $            & Bond albedo                           & -                        \\
	$ B_\lambda $      & Black body monochromatic intensity    & W m$ ^{-3} $ sr$ ^{-1} $ \\
	$ C $              & Constant in Eq \ref{eq:D-pV reln}     & km           \\
	$ C_1 $            & Constant in Eq \ref{eq: Bond albedo}  & -           \\
	$ C_2 $            & Constant in Eq \ref{eq: Bond albedo}  & -           \\
	$ c_s $            & Specific heat                         & J kg$ ^{-1} $ K$ ^{-1} $ \\
	$ D $              & Effective diameter                    & km                        \\
	$ F_{\nu}^{\rm filt} $  & In-band flux density             & Jy or W m$ ^{-2} $ Hz$ ^{-1} $ \\
	$ F_\nu $          & Flux density                          & Jy or W m$ ^{-2} $ Hz$ ^{-1} $ \\
	$ F_\nu^{\rm obs} $& Observed flux density                 & Jy or W m$ ^{-2} $ Hz$ ^{-1} $ \\
	$ F_\nu^{\rm model} $& Model flux density                  & Jy or W m$ ^{-2} $ Hz$ ^{-1} $ \\
	$ f $              & Filter transmission function          & -                        \\
	$ G $              & Phase function slope parameter        & -                        \\
	$ H_V $            & Absolute magnitude in V-band          & (mag)                    \\
	$ l_s $            & Diurnal thermal skin depth            & m                        \\
	$ N $              & Total number of data points           & -                        \\
	$ n $              & Number of free parameters             & -                        \\
	$ P $              & Rotational period                     & hour                     \\
	$ p_V $            & Geometric albedo in V-band            & -                        \\
	$ p_R $            & Geometric albedo in R-band            & -                        \\
	$ q $              & Phase integral                        & -                        \\
	$ r_h $            & Heliocentric distance                 & au or m                        \\
	$ \mathbf{S} $     & Rotational pole vector ($ \lambda_s $, $ \beta_s $) & ($ ^{\circ} $)  \\
	$ T $              & Temperature                           & K                        \\
	$ T_0 $            & Hypothetical subsolar temperature     & K                        \\
	$ t $              & Time                                  & s                        \\
	$ z $              & Depth (0 is surface)                  & m                        \\
	$ \alpha $         & Phase angle                           & ($ ^{\circ} $)           \\
	$ \beta_s $        & Ecliptic latitude of pole vector      & ($ ^{\circ} $)           \\
	$ \Gamma $         & Thermal inertia                       & SI $\equiv$ J m$ ^{-2} $ K$ ^{-1} $ s$ ^{-1/2} $\tablenotemark{a}     \\
	$ \Delta $         & Geocentric distance                   & au or m                        \\
	$ \Delta \lambda $ & Wavelength interval                   & $ \mu $m                        \\
	$ \Delta \nu $     & Frequency interval                    & Hz                       \\
	$ \epsilon_{\rm ran} $ & Random error of observation       & Jy or W m$ ^{-2} $ Hz$ ^{-1} $ \\
	$ \epsilon_{\rm sys} $ & Systematic error of observation   & Jy or W m$ ^{-2} $ Hz$ ^{-1} $ \\
	$ \varepsilon_h $  & Hemispherical emissivity (0.900)      & -                        \\
	$ \Theta $         & Thermal parameter                     & -                        \\
	$ \kappa $         & Thermal conductivity                  & W m$ ^{-1} $ K$ ^{-1} $  \\
	$ \lambda $        & Wavelength                            & $ \mu $m                        \\
	$ \lambda_s $      & Ecliptic longitude of pole vector     & ($ ^{\circ} $)           \\
	$ \rho $           & Mass density                          & kg m$ ^{-3} $            \\
	$ \chisq $         & Reduced chi-square statistic          & -                        \\
	$ \omega $         & Rotational angular velocity           & (rad) s$ ^{-1} $         \\
	\enddata
	\tablenotetext{a}{See footnote 1.}
	\tablecomments{Variables with hiphen(-) in third column are dimensionless. The units are basically given in SI format unless special units are dominantly used in this work. The units given in parentheses are dimensionless but are preferred to be explicitly written.}
	\tabletypesize{\small}
\end{deluxetable}

\section{Observations and Data Reduction} \label{sec:Observation}
The Japanese infrared satellite AKARI \citep{Murakami2007} was launched on 2006 February 21 UT, and its liquid helium cryogen boiled off on 2007 August 26 UT, 550 days after the launch (this cryogenic phase is called Phase 1, and 2). In the post-helium phase (Phase 3), the telescope and its scientific instruments were kept around 40 K by the mechanical cooler and only near-IR observations were carried out until 2010 February. 

This study is based on the spectroscopic data obtained by the Infrared Camera (IRC; \citealt{Onaka2007}) on board AKARI during both the cryogenic phase (Phase 2; for HR30) and the post-helium phase (Phase 3; for 4015WH). Our targets were observed mainly as part of the AKARI Mission Program ``Origin and Evolution of Solar System Objects'' (SOSOS). The observation log is summarized in Table \ref{tab: observation}.

\begin{deluxetable*}{ccccccccc} 
	\centering
	\tablecaption{Observational Circumstance. \label{tab: observation}} 
	\tablewidth{1.0\columnwidth}
	\tablehead{\colhead{Target} & \colhead{ID} & \colhead{Observed Time (UT)} & \colhead{Observation Mode} & \colhead{$ r_h $ (au)} & \colhead{$ \Delta $ (au)} & \colhead{$ \alpha $} & \colhead{Comment}} 
	\startdata
	4015WH  & 1521116-001 & 2009-11-18 13:57:26 & IRCZ4, b; Np (grism) & 1.05882 & 0.379283 & 69.0$ ^{\circ} $ & MP-SOSOS \\
	\hline
	& 1500821-001 & 2007-01-14 18:34:48 & IRC04, b; Np (grism) & 1.23927 & 0.756455 & 52.5$ ^{\circ} $ & MP-SOSOS \\
	& 1500820-001 & 2007-01-15 17:44:50 & IRC04, b; Np (grism) & 1.24132 & 0.754795 & 52.4$ ^{\circ} $ & MP-SOSOS \\
	HR30 & & 2007-01-14 08:54:10 & Survey Mode (S9W)    & 1.23846 & 0.757145 & 52.5$ ^{\circ} $ & $ 9.511 \pm 0.529 $ Jy\\
	& & 2007-01-14 15:31:17 & Survey Mode (S9W)    & 1.23902 & 0.756631 & 52.5$ ^{\circ} $ & $ 8.288 \pm 0.461 $ Jy\\
	& & 2007-01-13 08:04:38 & Survey Mode (L18W)   & 1.23650 & 0.759207 & 52.7$ ^{\circ} $ & $ 12.567 \pm 0.838 $ Jy\\
	& & 2007-01-13 09:45:56 & Survey Mode (L18W)   & 1.23663 & 0.759063 & 52.6$ ^{\circ} $ & $ 12.145 \pm 0.810 $ Jy\\
	\enddata
	\tablecomments{For survey mode data, the observed flux densities and their 1-$\sigma $ uncertainties are given in Jy unit.}
\end{deluxetable*}

The IRC has a spectroscopic capability in both the grism mode and the prism mode with the AKARI Astronomical Observation Template (AOT) IRC04 (Phase 2) or IRCZ4 (Phase 3) \citep{Onaka2007,Onaka2010,Ohyama2007}. The grism mode and the prism mode can cover the wavelength range from 2.5 to 5 $ \mu $m with the spectral resolution of $R \sim 100$, and from 1.7 to 5.4 $ \mu $m with $R \sim $ 20--40, respectively. Within a single AOT operation lasting about 10 minutes, 8 or 9 spectroscopic images with either the grism or the prism, as well as a direct image (called a reference image) through a broadband filter centered at 3.1 $ \mu $m, are taken. The effective exposure time for each frame is 44.41 sec. The targets were put on the 1 arcminute $ \times $ 1 arcminute aperture mask (the ``Np'' window; see Fig.3 in \citealt{Onaka2007}) to minimize the contamination from nearby objects (e.g., \citealt{Ootsubo2012a}). In this paper, we concentrate on the data taken using the grism mode. Three observations (ID of 1521116-001 for 4015WH, 1500820-001 and 1500821-001 for HR30) are used in total for the analyses below.

The raw data were basically reduced with the IDL-based software package, the IRC Spectroscopy Toolkit for the Phase 3 data, version 20150331 (\citealt{Ohyama2007}; also see the IRC Data User Manual\footnote{http://www.ir.isas.jaxa.jp/AKARI/Observation/support/IRC/}). The standard array image processing, such as dark subtraction, linearity correction, flat-field correction, and various image anomaly corrections, were first performed with the toolkit. Then the two-dimensional spectral images of the objects were extracted with the toolkit.

Since AKARI did not have a tracking mode for moving objects, the resultant two-dimensional spectrum with the toolkit was blurred because of the motion of the object. The movements of the objects during the observations of the frames were calculated, and individual data frames were shift-and-added \citep{Ootsubo2012a}. Using the combination of the shift-and-add with the movement of the object and the 3-sigma clipping methods, the effects of bad or hot pixels can be reduced, although the number of bad or hot pixels in Phase 3 was increased compared with Phase 2.

Finally, one-dimensional spectrum was extracted from the two-dimensional image by summing signals within 7 pixels (about 10.5 arcsecond) in the spatial direction from the source's central position. The background flux was estimated from the adjacent region of the target and subtracted.

It is reported that the sensitivity decreased about 10 \% at maximum in Phase 3 compared with that in Phase 2, which depends on the IRC detector temperature \citep{Onaka2010}. This can be corrected by using a formula with the detector temperature (Baba et al., in prep). For the observation ID of 1521116-001, the detector temperature was recorded as 45.128 K and the correction factor is given as 0.954238, while this absolute calibration was not carried out in this work. Note that in the Phase 2 observations (ID of 1500820-001 and 1500821-001), the temperature was kept 10.45 K and the observed fluxes are no need to be corrected. The systematic error is, however, small compared to the uncertainties in size of the targets (see Section \ref{sec:Results}), so we ignored the error throughout this work.

In the spectroscopic analyses, the flux and the wavelength accuracy strongly depend on the determination accuracy of the wavelength zero point, which was done on the reference image. Its accuracy is estimated to be at worst 1 pixel in our analysis. Thus the flux and the wavelength uncertainties were estimated by calculating how much the spectrum changes when the wavelength zero point shifted by $ \pm $ 1 pixel \citep{Shimonishi2013}.

We thoroughly have confirmed that the comet HR30 showed no visible cometary activity near the spectroscopic observations, by investigating the survey mode data (See Fig 6. of \citealt{Usui2011a}). From visual inspection, we could not detect any systematic elongation of HR30 with respect to the antisolar direction or the velocity vector; HR30 was elongated into random directions. Furthermore, the point spread functions of HR30 did not differ from those of bright stars in images from 2006-11-23, 2007-01-13, and 2007-03-29 (all UT). 


\section{Methodology}   \label{sec:Methodology}

\subsection{The Thermophysical Model} \label{subsec:TPM}
In this subsection, we describe the physical assumptions of our TPM and then introduced some relations between the model parameters that we implemented in the code.

\subsubsection{Assumptions}
A simple TPM generally has the following assumptions:
\begin{itemize}
	\setlength\itemsep{0em}
	\item non-tumbler with constant rotational period,
	\item spherical shape,
	\item one-dimensional heat conduction,
	\item constant thermophysical properties ($ \kappa $, $ \rho $, and $ c_s $, and hence $ \Gamma $ ) over the time, space, and temperature range of interest\footnote{The temperature dependency of $ \Gamma $  may have certain effect of roughly $ \Gamma \propto r_h^{-3/4} $ \citep{Mueller2010}. The uncertainty from this effect is smaller than the uncertainty from the model fitting at least for our targets.}, 
	\item temperature remains constant deeper than certain region,
	\item the standard $ H $, $ G $ magnitude system (Section \ref{subsec: parameter relations})
	\item constant thermal hemispherical emissivity ($ \varepsilon_h  = 0.9$), i.e., ignore emissivity dependence on direction and wavelength, so constant Bond albedo,
	\item rotational period much shorter than orbital period (seasonal effect is ignored),
	\item and the effective diameter determined in the V-band.
\end{itemize}
We further assume zero surface roughness. Neither chemical energy, such as sublimation or latent heat, nor external energy, such as cosmic rays or impacts, are considered throughout this study. Also the surface is assumed to be optically thick enough that neither the influx of solar energy from the top nor (thermal) radiation from the bottom can penetrate to the deeper region, and the thermal emissivity does not have to be low as it is in radio observations.

\subsubsection{Thermal Conduction Modeling}
Under these assumptions, the 1-dimensional heat conduction equation for the temperature $ T = T(z, t) $ is given as
\begin{equation} \label{eq: heat eq}
\rho c_s \frac{\partial T}{\partial t} = \kappa \frac{\partial^2 T}{\partial z^2} 
\end{equation}
with two boundary conditions. The first is
\begin{equation}
\left . \frac{\partial T}{\partial z} \right | _{z \gg l_s} = 0 ~,
\end{equation}
where $ l_s $ is the thermal skin depth (see Eq \ref{eq: thermal skin depth}). For the second boundary condition, a balance between ``absorbed solar energy plus conduction from beneath'' and ``emitted thermal radiation'' is used for the surface ($ z=0 $).The practical implementation of $ z \gg l_s $ is discussed in Subsection \ref{subsec: code implementation}.

The equilibrium gray body (with hemispherical emissivity $ \varepsilon_h $) temperature at constant $ r_h $ is \citep{Harris1998}:
\begin{equation} \label{eq: T_0}
T_0 = \sqrt[4]{\frac{(1-A_B)}{ \varepsilon_h \sigma_{\rm SB} } \frac{L_{\sun}}{4\pi r_h^2} } ~.
\end{equation}
$ T_0 $ is not necessarily the same as the real \emph{subsolar} temperature if the object is rotating and has non-zero $ \Gamma $, so we avoid using $ T_{SS} $, which might be misleading but is nevertheless widely used. 

If the seasonal effect is negligible, the thermal skin depth is only affected by diurnal motion and can be defined as (e.g., \citealt{Spencer1989}):
\begin{equation} \label{eq: thermal skin depth}
l_s = \sqrt{\frac{\kappa}{\rho c_s \omega}} ~.
\end{equation}
Usually $ l_s $ is on the scale of $ 10^{-3} $ to $ 10^{-1} $ meters, which is much smaller than the computational resolution of the TPM in the horizontal direction (see Subsection \ref{subsec: code implementation}). The vertical (depth) resolution is usually on the order of $ 0.1 l_s $; thus we are justified in using the 1-D equation (Eq \ref{eq: heat eq}) rather than solving the more complicated 3-dimensional heat conduction equation.

Under the aforementioned assumptions, the thermal emission profile is dependent on a dimensionless parameter, called the thermal parameter \citep{Spencer1989}:
\begin{equation} \label{eq: Theta-TI reln}
\Theta \equiv \frac{\Gamma \sqrt{\omega}}{\varepsilon_h \sigma_{\rm SB} T_0 ^3} ~,
\end{equation}
for a given spin orientation.

\subsubsection{Parameter Relations} \label{subsec: parameter relations}
To model an asteroid in thermal equilibrium, it is inevitable to use the bolometric Bond albedo, which is not generally obtained by observation. It is, however, possible to approximate it using the so-called standard $ H $, $ G $ magnitude system \citep{Bowell1989}. In this system, the phase function of a general asteroid is empirically determined, and the Bond albedo for a Lambertian sphere with phase correction is written as
\begin{equation} \label{eq: Bond albedo}
A_B = A_B(G, p_V) = q(G) p_V = (C_1 + C_2 G) p_V ~,
\end{equation}
Although \cite{Bowell1989} derived $ (C_1, C_2) = (0.290,~ 0.684) $, \cite{Myhrvold2016a} found that the best fitting function had $ (C_1, C_2) = (0.286,~ 0.656) $. We confirmed the latter result and used it throughout this study. The $ G $ and $ p_V $ values are relatively easier to obtain than the Bond albedo, so Eq \ref{eq: Bond albedo} is a useful tool for modeling asteroid thermal emission.

Once we assume a spherical model asteroid, we need a representative diameter, namely, the \textit{effective} diameter. The effective diameter is defined as the diameter of a Lambertian sphere with a geometric albedo that emits the same amount of total flux (or absolute magnitude) at the given wavelength range as the observed asteroid, at perfect opposition. Then, $ H_V $ and $ p_V $ are related to the effective diameter as (detailed derivation is given in \citealt{Pravec2007}):
\begin{equation}\label{eq:D-pV reln}
D = D(p_V, H_V) = \frac{C}{\sqrt{p_V}} 10^{-H_V/5}~, 
\end{equation}
where $ C = (2{\rm au}) \times 10^{V_{\sun}/5}$. This relation reduces one of the free parameters, either the effective diameter (in the V-band, rigorously speaking) or $ p_V $. Following \cite{Fowler1992}, we used $ C=1329 $ km.

These relations simplify the model because the thermal parameter (Eq \ref{eq: Theta-TI reln}) can be parameterized as
\begin{equation}
\Theta = \Theta (\Gamma, \omega, T_0) = \Theta(\Gamma, \omega, G, D, r_h) ~.
\end{equation}
The parameter $ G $ has a negligible effect (see next section), $ r_h $ is known a priori from the ephemeris, and $ \omega $ have already been obtained from previous studies \citep{Urakawa2011,Harada2009}. Thus, we considered $ \Gamma $ and $ D $ as free parameters.

\subsection{Model Parameters} \label{subsec: model parameters}

Some of the parameters are known from previous studies, and the known values are listed in Table \ref{tab: known parameters}. For HR30, we accepted a nominal absolute magnitude, $ H_R = 11.99 \pm 0.01 $, which can be converted to $ H_V = 12.09 \pm 0.013 $, obtained from \cite{Hicks2007}, as the true value. Their $ H_R $ was obtained from the photometric magnitude of the nucleus in the R-band ($R = 15.69 \pm 0.01$) at $ \alpha \sim 17^{\circ} $ by assuming $ G = 0.0 $ and $ p_R = 0.05 $. 

\begin{deluxetable}{lll} 
	\tablecaption{Known parameters \label{tab: known parameters}}
	\tablehead{ \colhead{Quantities} & \colhead{4015WH} & \colhead{HR30} }
	\startdata
	$ G $    
	& $ 0.07 \pm 0.03 $\tablenotemark{a} 
	& - \\
	$ H_V $	 
	& $ 15.90 \pm 0.10 $\tablenotemark{a}
	& $ 12.09 \pm 0.013 $\tablenotemark{d}\\
	$ P $ [hours]
	& 7.15\tablenotemark{b}
	& 73.2\tablenotemark{e, f} \\
	$ \textbf{S} (\lambda_s, \beta_s) $	
	& $ (330^{\circ}, -27^{\circ})$\tablenotemark{b} 
	& -\\
	$ p_R $  
	& $ 0.055 \pm 0.012 $\tablenotemark{a} 
	& -  \\
	$ p_V $         
	& $ 0.059 \pm 0.011 $\tablenotemark{c}
	& - \\	  
	$ D $ [km]
	& $ 3.63\pm 0.56 $\tablenotemark{a}
	& 22\tablenotemark{d}\\
	& $ 3.46 \pm 0.32 $\tablenotemark{c} & \\
	$ \Gamma $ [SI]
	& $ > 60 $\tablenotemark{c}
	& - \\
	\enddata
	\tablenotetext{a}{\cite{Ishiguro2011}. Using the standard $ H $, $ G $ magnitude system \citep{Bowell1989}.}
	\tablenotetext{b}{\cite{Urakawa2011}. The non-tumbling solution.} 
	\tablenotetext{c}{\cite{Licandro2009}. Using NEATM \citep{Harris1998}.}
	\tablenotetext{d}{\cite{Hicks2007}. Assumed $ G=0.0 $ and $ p_R = p_V = 0.05 $.}
	\tablenotetext{e}{\cite{Harada2009}.} 
	\tablenotetext{f}{\cite{Galad2008} obtained 68, 70.7, or 73 h as possible rotational periods. We adopted 73.2 h from \cite{Harada2009}, as it coincides well with this published data.}
\end{deluxetable}

The resulting temperature distributions are very robust against changes in $ G $, so we can safely fix this value. To be more rigorous, the fractional change in $ T_0 $ can be obtained by differentiating Eq \ref{eq: T_0} by $ A_B $ and substituting Eq \ref{eq: Bond albedo}:
\begin{equation}\label{eq: G-error validation}
\frac{\delta T_0}{T_0} \approx \frac{C_2 G p_V}{4(1-A_B)} \frac{\delta G}{G} = 8.4 \times 10^{-3} \frac{\delta G}{G} ~.
\end{equation}
The second equality is calculated using nominal values, $ p_V = 0.05 $ and $ G = 0.15 $. Various values of $ G $ affect the resultant flux density by a few percent ($ 0.0 \le G \le 0.15$). Even when $ G $ was increased up to 0.5, the flux density differed by only $ \lesssim 5\% $ compared to that of $ G = 0.0 $ case. This is even smaller than the systematic error of the AKARI observation; therefore, we fixed $ G $ as 0.07 and 0.15 for 4015WH and HR30, respectively.

Two observation epochs for HR30 were made at very similar times using grism (Table \ref{tab: observation}). We confirmed that the difference in model calculations using two different ephemerides was not significant (flux density fractional difference $ \lesssim 3\% $ at all wavelengths), so we regarded those two datasets as a single epoch, viz., 2007-01-14 18:34:48 (UT). Finally, the rotational period, $ P $, was also fixed since it only appears in Eq \ref{eq: Theta-TI reln} to determine $ \Theta $. If the $ P $ value is updated to $ P' $, we can simply multiply $ \sqrt{P'/P} $ by our found $ \Gamma $ value. Since the pole orientation is known for 4015WH (Table \ref{tab: known parameters}), we fixed this value, but it was set as a free parameter for HR30.

In summary, we fixed $ G=0.07 $, $ H_V = 15.90 $, $ P = 7.15 $ h, and $ \textbf{S} = (330^{\circ}, -27^{\circ}) $ for 4015WH, and $ G=0.15 $, $ H_V = 12.09 $, and $ P = 73.2 $ h for HR30. Then we are left with two parameters for 4015WH ($ D $, $ \Gamma $), and two more (spin orientation) for HR30. The effective diameter $ D $ is derived by using Eq \ref{eq:D-pV reln} with $ p_V $ ranging from 0.02 to 0.07 with 0.005 interval, while $ \Gamma = \{ 1,~ 10,~ 100,~ 250,~ 500,~ 750,~ 1000,~ 2000,~ 3000 \}$ SI is used. For HR30, $ D $ is derived by varying $ p_V $ form 0.02 to 0.06 with 0.005 interval. We further considered the spin orientation of HR30 in the ranges of the longitude $ \lambda_s \in [0^{\circ}, 360^{\circ}] $ and the latitude $ \beta_s \in [-90^{\circ}, +90^{\circ}] $ with 15$ ^{\circ} $ and 10$ ^{\circ} $ intervals, respectively. The fixed parameter values and the searched parameter space domains are summarized in Table \ref{tab: parameter space}.

\begin{deluxetable}{lll} 
	\tablecaption{Fixed and searched parameter space. \label{tab: parameter space}}
	\tablehead{ \colhead{Quantities} & \colhead{4015WH} & \colhead{HR30} }
	\startdata
	$ G $    
	& \textit{0.07}
	& \textit{0.15} \\
	$ H_V $	 
	& \textit{15.90}
	& \textit{12.09} \\
	$ P $ [hours]
	& \textit{7.15}
	& \textit{73.2} \\
	$ \textbf{S} = (\lambda_s, \beta_s) $	
	& $ \mathit{(330^{\circ}, -27^{\circ})} $
	& all directions \\
	$ D $ [km]
	& 3.3--6.2 
	& 19--29\\
	$ p_V $  
	& 0.02--0.07 
	& 0.02--0.06 \\
	$ \Gamma $ [SI]
	& 1--3000
	& 1--3000\\
	\enddata
	\tablecomments{Values in italic are fixed parameters. See the text for detailed searching grid.}
\end{deluxetable}

\subsection{Code Implementation} \label{subsec: code implementation}
We applied a strategy similar to that of \cite{Mueller2007} to solve Eq \ref{eq: heat eq}. We set the goal accuracy (the fractional temperature difference on the surface after one full rotation) to $ 10^{-5} $ after minimum iteration 50, and the resolutions of the model of 1$^{\circ}$ in longitude, 4$^{\circ}$ in latitude, and 0.25 $ l_s $ in depth. For the depth, the deepest depth is set to be $ 8.0 l_s $. Increasing this depth to $ 10.0 l_s $ affected the final equilibrium temperature only $ \lesssim 0.1 $\%. The flux density, as well as the temperatures for each longitude, latitude, and depth slab, are saved.

Once we obtained the flux density for each wavelength, we calculated the reduced chi-square statistic for each model with respect to the corresponding observational data, which is defined as
\begin{equation}\label{eq: red chi^2 defn}
\chi^2_{\rm red} \equiv \frac{1}{N-n} \sum_{i=1}^{N} \frac{( F_\nu^{\rm obs}(\lambda_i) - F_\nu^{\rm model}(\lambda_i) )^2 }{\epsilon_{\rm ran}^2} ~,
\end{equation}
where $ N $ and $ n $ are the number of observations and free parameters, so $ N-n $ is the total degrees of freedom. Generally the systematic error is not included in the denominator since it is not a Gaussian noise. The dummy variable $ i $ indicates the wavelength bin. Only the data with $ 3.5 ~\mu \text{m} \le \lambda \le 4.8 ~\mu \text{m} $ are used for this calculation since the reflected sunlight component dominates thermal radiation at shorter wavelengths. At longer wavelengths, there is a calibration error in the measured flux density \citep{Baba2016} and not yet corrected in this data reduction, so we set the upper limit by visual inspection. 

Our approach slightly differs from some most recent TPMs, e.g., in \cite{Mueller2007} and \cite{Muller2017}, in a sense that the so-called \emph{scale factor} is not multiplied to $ F_\nu^{\rm model} $ of Eq \ref{eq: red chi^2 defn}. Instead of finding the best fit scale factor, we employed brute-force parameter space searching. Thus, we re-calculated the temperature map on the asteroidal surface for each parameter pair, and compared the model flux with the observed data. This is computationally expensive but useful when there are small number of free parameters as in our case.

Following \cite{Hanus2015}, we adopt the criterion of $ \chisq < \chi^2_{\rm red, min} + \sqrt{\frac{2}{N-n}} $ to estimate the confidence interval of the free parameters (e.g., Chapter 15 of \citealt{Press2007} and \citealt{Hanus2015}). We also checked whether the minimum reduced chi-square statistic, $ \chi^2_{\rm red, min} $, is close to 1.

We used the survey mode data for HR30 (See Fig 6. of \citealt{Usui2011a}) as criteria to reject models that do not give appropriate in-band flux\footnote{Another possibility is to introduce the maximum compatible estimators for weighting function \citep{Kaasalainen2011}.}. The corresponding in-band flux is obtained in flux density units (e.g., Jy) using the following formula:
\begin{equation}\label{eq: filter conv}
F_{\nu}^{\rm filt} = \frac{\sum_{i} F_{\nu}^{\rm model} (\lambda_i) f(\nu_i) \Delta \nu_i}{\sum_{i} f(\nu_i) \Delta \nu_i} ~.
\end{equation}
The in-band model flux can be calculated by substituting $ f(\nu_i) \Delta \nu_i = c f(\lambda_i) \Delta \lambda_i / \lambda_i^2$. It is used only to reject models and is \emph{not} used in calculating $ \chisq $.

\section{Results}   \label{sec:Results}
The best-fit model parameters for the two targets are described here for each target. 


\subsection{4015WH} 
\begin{figure} 
	\centering
	\includegraphics[width=\linewidth]{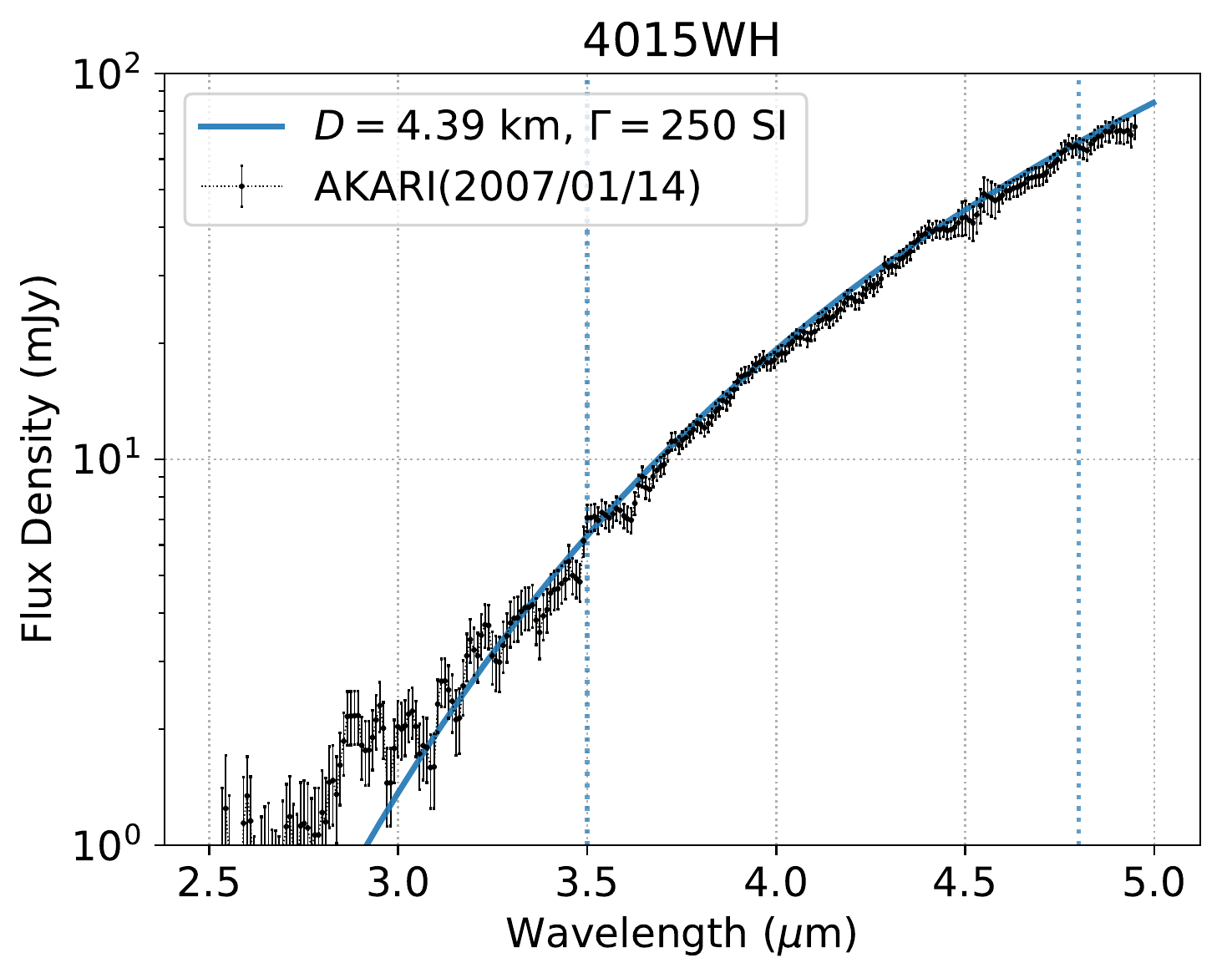}
	\caption{The spectral energy distribution plot of the best fit model in solid lines with label. The AKARI observations are overplotted with random noise ($ \epsilon_{\rm ran} $) as the error bars. Vertical thick dotted lines indicate the 3.5 and 4.8 $ \mu $m wavelengths, which are the boundaries of the wavelengths we used for the chi-square minimization. \label{fig: 4015 flux}}
	\label{fig:4015WH_flux}
\end{figure}

\begin{figure} 
	\centering
	\includegraphics[width=1\linewidth]{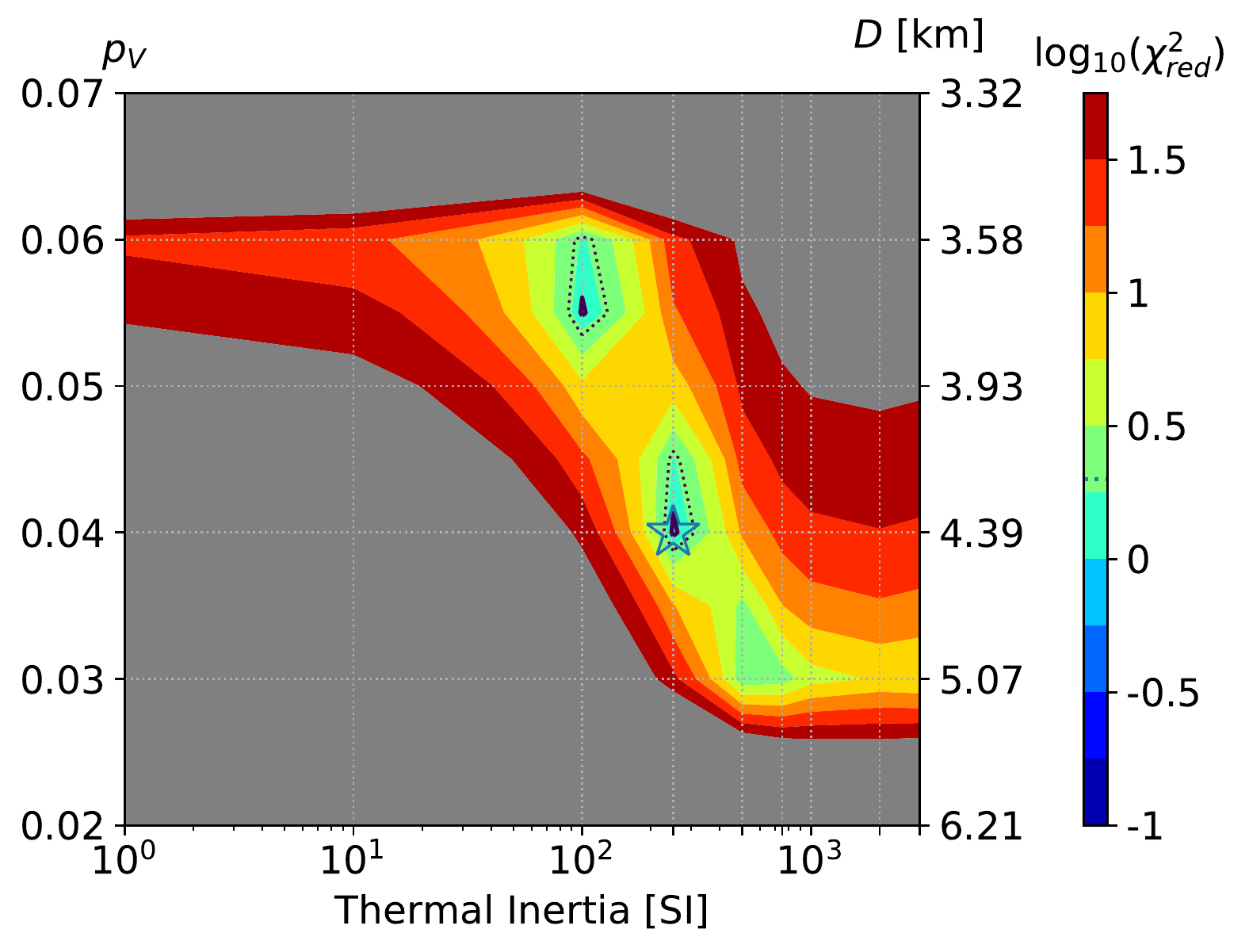}
	\caption{The plot of $ \log_{10}(\chisq) $ for 4015WH in the parameter space. The gray region is where $ \chisq > 100 $. The star marker shows the location where the minimum $ \chisq $ occurred ($ D = 4.4 $ km, $ \Gamma = 250$ SI), and the tiny solid line contour is the confidence interval (see Subsection \ref{subsec: code implementation}). Dotted contours, which represent $ \chisq = 2.0 $, are also shown to guide the eyes. \label{fig: 4015 chisq} }
	\label{fig:4015WH_chisq}
\end{figure}

The smallest $ \chisq $ appeared at $ D = 4.4 $ km and $ \Gamma = 250 $ SI with $ \chisq = 1.157 $. The model is plotted together with AKARI observational data (Fig \ref{fig: 4015 flux}). We also plotted the $ \log_{10} \chisq $ contour map in Fig \ref{fig: 4015 chisq}. We obtained $ D = 3.74$--4.39 km and $ p_V = 0.04 $--0.055 with $ \Gamma = 100 $--250 SI for the confidence interval. Note that the systematic deviation of AKARI data from the model at shorter wavelength in Fig \ref{fig: 4015 flux} is because of the reflected sunlight component.

\subsection{HR30} 
For HR30, we compared more than 40,000 models in the 4-dimensional parameter space spanned by the pole orientation, as well as $ D $ ($ p_V $) and $ \Gamma $. We accepted models which (1) meet the $ \chisq $ criterion of Subsection \ref{subsec: code implementation} and (2) had the in-band flux ($F_\nu^{\rm filt}$; Eq \ref{eq: filter conv}) within the 3-$ \sigma $ range of S9W and L18W data.

After applying the criteria, we were left with only 23 models (for comparison, $ \chisq < 2 $ left 201 models). The two minimum $ \chisq $ models are: pole $ (330^{\circ}, 10^{\circ}) $ with $ D = 27 $ km, $ p_V = 0.035 $ and $ \Gamma = 250 $ SI ($ \chisq = 1.181 $) and $ (225^{\circ}, 30^{\circ}) $ with the same $ D $, $ p_V $, $ \Gamma $, and $ \chisq $ values.



\begin{figure} 
	\centering
	\includegraphics[width=\linewidth]{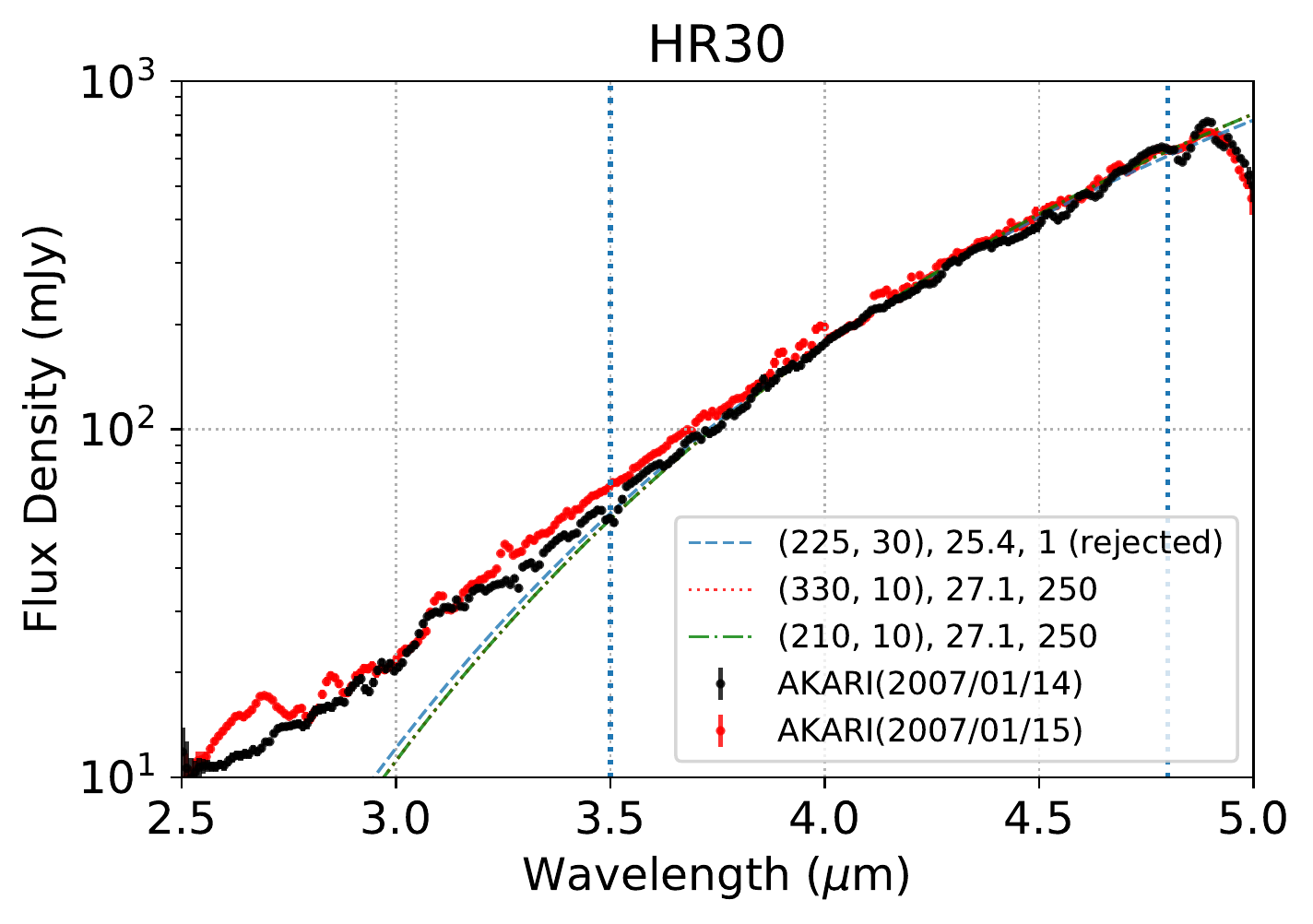}
	\caption{Three good-fitting models are plotted with the AKARI observation data. The labels follow ``$ (\lambda_s, \beta_s) $, $ D $ (km), $ \Gamma $ (SI)'' notation. The first model is rejected by the broadband criteria (see text and Fig \ref{fig:HR30_broad}). Vertical thick dotted lines indicate the 3.5 and 4.8 $ \mu $m wavelengths, which are the boundary wavelengths we used for the chi-square minimization. The models are almost indistinguishable in the figure.}
	\label{fig:HR30_flux}
\end{figure}

\begin{figure}
	\centering
	\includegraphics[width=\linewidth]{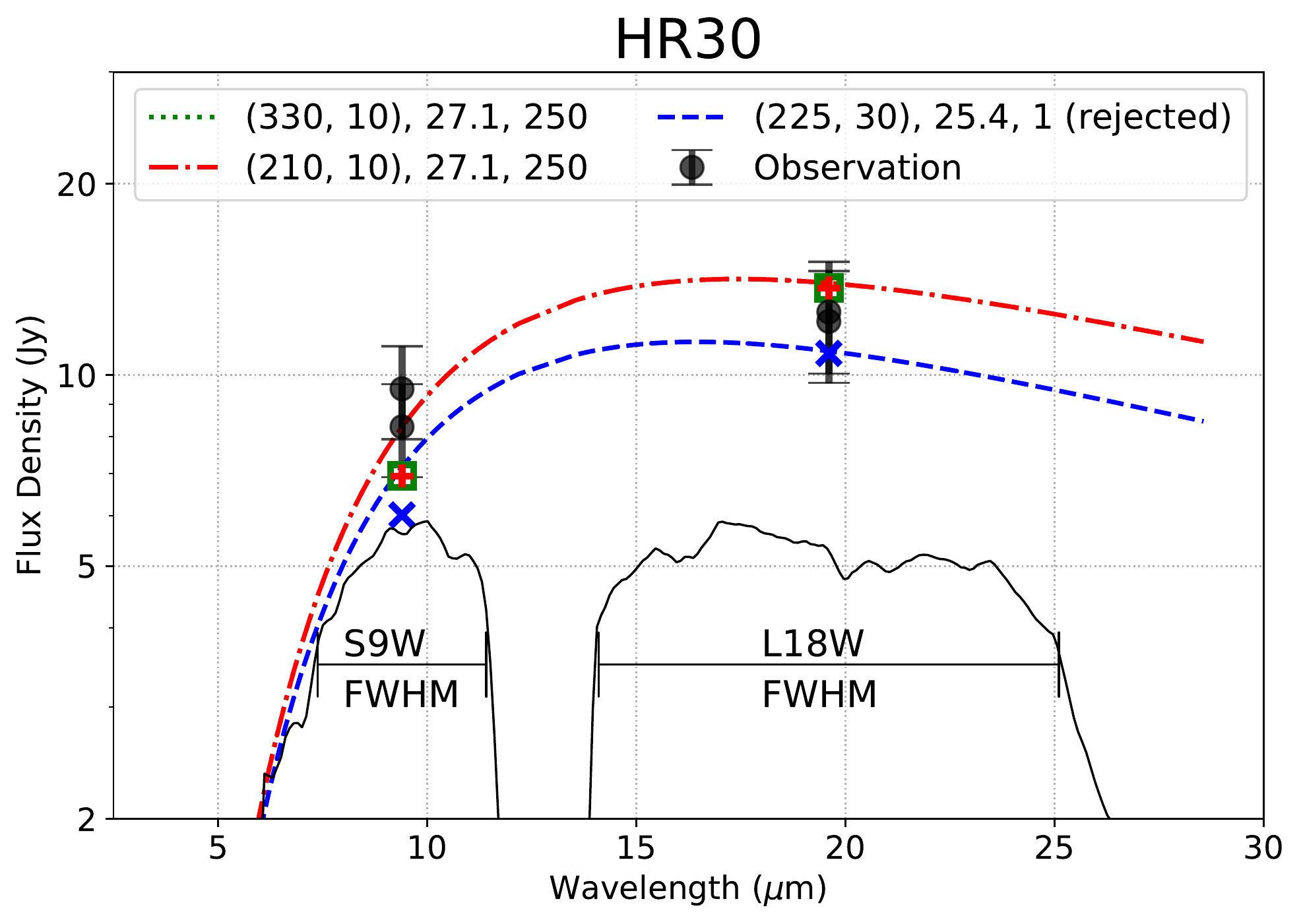}
	\caption{The survey mode observations and model calculations are shown with the same legends as \ref{fig:HR30_flux}. Thin solid lines with text show the profiles of the S9W and L18W filters in logarithmic scale for comparison with FWHMs. The survey mode data (see Table \ref{tab: observation}) are plotted as filled circles centered at the central wavelengths with vertical 3-$ \sigma $ error bars of each observation. The markers show the in-band fluxes from each model of the same color. The blue dashed model is rejected because it is out of range of the S9W data.}
	\label{fig:HR30_broad}
\end{figure}

These two good-fitting models of HR30 are compared with the observation data in Fig \ref{fig:HR30_flux}, and the in-band flux data are compared in Fig \ref{fig:HR30_broad}. For comparison, we plotted one of the rejected models (labeled as ``rejected''). It had low $ \chisq $ for grism data but failed in reproducing the survey mode data. In Fig \ref{fig:HR30_flux}, the observed fluxes deviate from the model much more than 4015WH at shorter wavelength. This is due to the stronger reflected light component from HR30.

\begin{figure*}
	\centering
	\includegraphics[width=\linewidth]{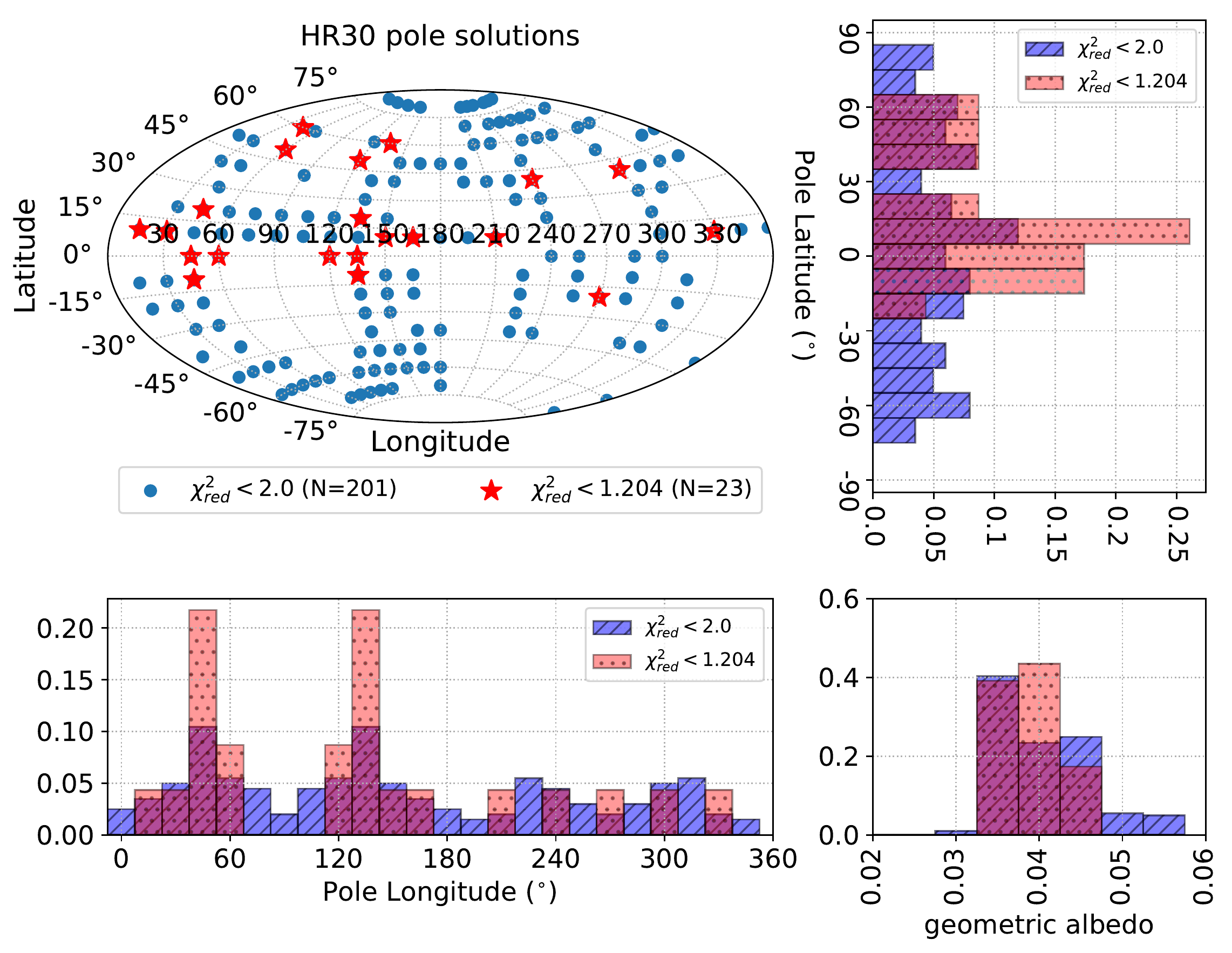}
	\caption{Top left: Plot showing the models for HR30 that accurately reproduced the observations. A marker is shown if at least one of the $ (D, \Gamma) $ pairs had $ \chisq < 2 $ (blue circles) or $ \chisq < \chi^2_{\rm red, min} + \sqrt{2/(N-n)} = 1.204$ (red stars). N in the legend is the number of models that meet the criteria. Only those that met the survey mode observation criteria are used. Top right and bottom left: Histogram showing the fraction of models with respect to the pole latitude and longitude (ecliptic), respectively. Two types of bars represent $ \chisq < 2 $ (blue dashed) and $ \chisq < 1.204 $ (red dotted). Bottom right: Histogram showing the fraction of models with respect to the geometric albedo. The bars are the same as in the aforementioned histograms.}
	\label{fig:HR30_Pole_Projection}
\end{figure*}

Fig \ref{fig:HR30_Pole_Projection}  shows the pole orientations of the acceptable models for HR30. The spin latitude of HR30 is likely to be near the ecliptic plane ($ \beta_s \sim 0 $), although some extreme cases, such as $ \beta_s \sim 60^{\circ} $, are not rejected. Taking the solid angle effect into account, i.e., weight with solid angle to each model, the fraction of models is more concentrated to $ \beta_s \sim 0 $. The longitude is uncertain, yet it is probabilistically more likely to be found at $ \lambda_s \sim 45^{\circ} $ or $ \lambda_s \sim 135^{\circ} $. The geometric albedo, and thus the size, is strongly concentrated at $ p_V \sim 0.04 $ ($ D \sim 25$ km). None of the models other than $ 23.9 ~{\rm km} \le D \le 27.1 ~{\rm km} $ could be accepted. The $ \Gamma $ values for HR30 are distributed from 250 to 1,000 SI with quite uniform frequency. 

If we apply the same strategy to obtain $ \chisq $ from the UT 2007-01-15 data, however, none of the accepted models reproduce the in-band fluxes of the survey mode within the 3-sigma range. We discuss about this in Subsection \ref{subsec: error sources}.

\subsection{Size and Thermal Inertia}

\begin{figure}
	\centering
	\includegraphics[width=1\linewidth]{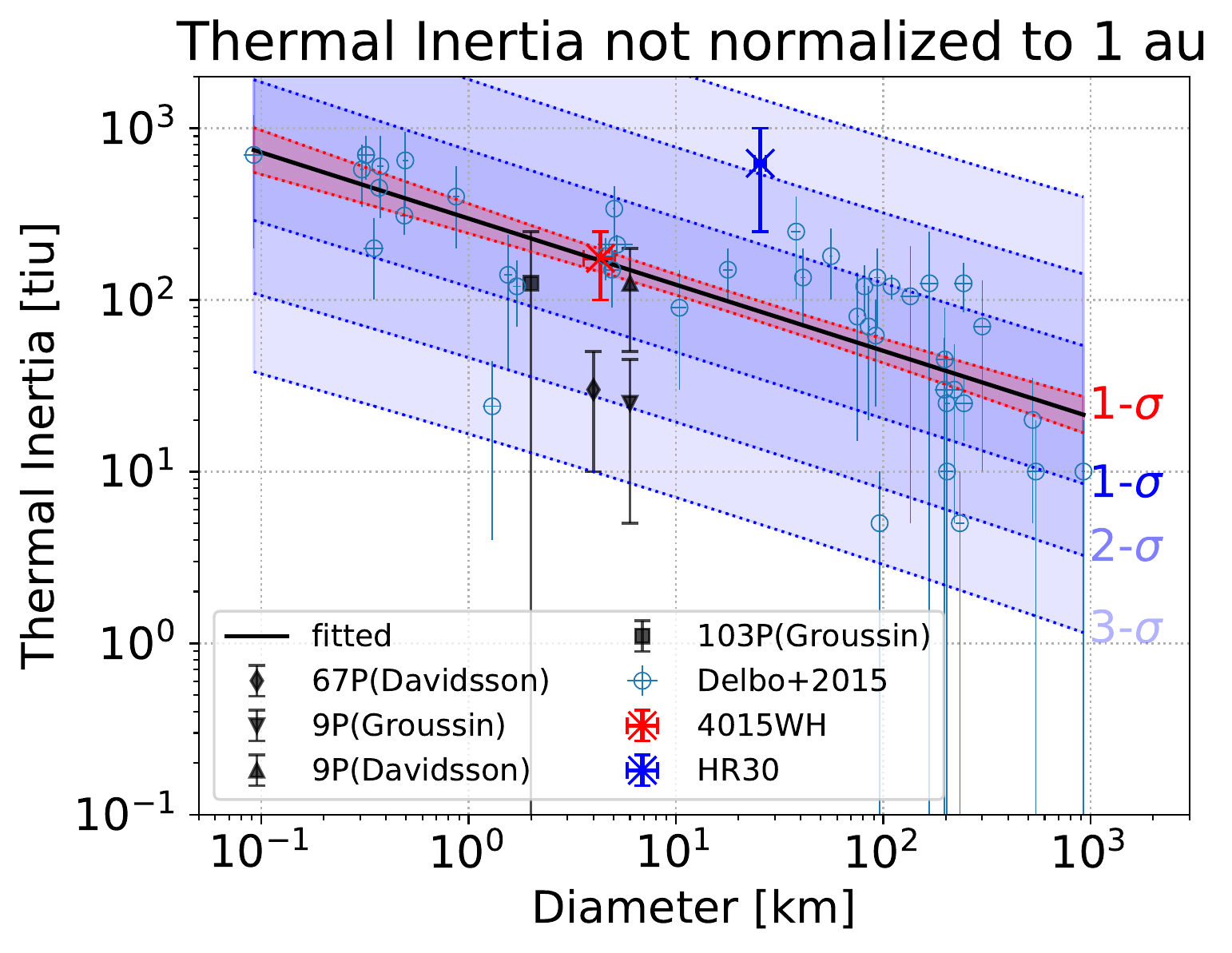}
	\caption{The $ D $--$ \Gamma $ distribution of asteroids and comets with the targets investigated in this study. The blue circles, with black best-fit line, are imported from \cite{Delbo2015a} with $ \Gamma $ values not normalized to 1 au. Some space mission target results are shown in black (see text). The red shaded region and three blue shaded regions show the confidence interval and prediction intervals correspond to the indicated significance level, respectively (see text). The uncertainties in the effective diameters are shown except for black markers (in situ observations).}
	\label{fig:TIvsD}
\end{figure}

We plotted the derived thermal inertias of 4015WH and HR30 with respect to the diameter (Fig \ref{fig:TIvsD}). For comparison, we imported the $D$--$\Gamma$ data from \cite{Delbo2015a}, excluding Jupiter Trojans, Centaurs, and TNOs from the original list. In addition to the asteroids, some of the comets that have been visited via spacecrafts and modeled in detail are also shown. They are 9P/Tempel 1 ($ D \sim 6 $ km and $ \Gamma$ $ \lesssim $ 45 SI or 50--200 SI; $ \Gamma$ from \citealt{Groussin2013} and \citealt{Davidsson2013}, respectively), 103P/Hartley 2 ($ D \sim $ 2 km and $ \Gamma$ $ \lesssim 250 $ SI; $ \Gamma$ from \citealt{Groussin2013}), and 67P/Churyumov--Gerasimenko ($ D \sim $ 4 km and $ \Gamma$ $ \lesssim 50 $ SI; $ \Gamma$ from \citealt{Gulkis2015a} and \citealt{Shi2016}). The diameter uncertainties of these space mission targets are not considered. 

To see the membership of our targets and the visited comets from asteroidal distribution, we did linear regression to asteroids. Then the 1-, 2-, and 3-$ \sigma $ prediction intervals (i.e., significance level of 0.6827, 0.9545, and 0.9973, respectively) are calculated and shown as blue shaded regions. The red shaded region shows the 1-$ \sigma $ confidence interval of fitting line. Although HR30 is slightly out of 2-$ \sigma $ prediction interval, so do few asteroids. All the small bodies reasonably lie within the 3-$ \sigma $ prediction interval. Now it is clear that the two of our targets are \emph{not} outside of the asteroidal trend. Therefore, we conjecture that comet-like objects may show similar, i.e., indistinguishable, trend compared to usual asteroids, on the $ D $--$ \Gamma $ plane.

\section{Discussion}   \label{sec:Discussion}
Our TPM succeeded in deriving the sizes, albedos, and thermal inertias for two objects. In addition, we estimated the pole orientation of HR30. In this section, we discuss limitations of our model, possible sources of uncertainty in the results, and the implications of the results.

\subsection{Shape and Roughness}
The limited amount of observational data limited the number of free parameters. These parameters, or the limitations, include the shape and surface roughness. The importance of the shape and roughness were carefully studied by \cite{Hanus2015} and \cite{Davidsson2015}, respectively. In this subsection, we justify the use of the smooth spherical model, i.e., excluding the shape and surface roughness effects.

Even if the target has an irregular/elongated shape and has varying roughness, the effective diameter (or the volume equivalent diameter) and thus the geometric albedo remain relatively constant. An example is given in Table 4 of \cite{Hanus2015}: Different diameter values were derived for each target by varying the shape models and roughness parameters, and all the obtained values were consistent within 1-sigma uncertainty in all cases, unless the uncertainty is not given. 

In the case of (25143) Itokawa, \cite{Muller2014} carefully compared a spherical TPM with light curve inversion and an in situ shape model, and they found three models produce consistent results. They found that a priori knowledge of the spin vector and rotational period can affect the reliability of TPM results, although a spherical model may give consistent results with more advanced techniques. For 4015WH, we already know the rotational period and pole orientation with certainty from previous studies (Table \ref{tab: known parameters}). The parameters ($ D $, $ p_V $, and $ \Gamma $) we derived are indeed consistent with the previous studies and the corresponding taxonomic type (see Subsection \ref{subsec: Implications}).

Furthermore, at large phase angles, as in our observation ($ \alpha > 50^{\circ} $; see Table \ref{tab: observation}), the effect of roughness might weaken especially at wavelengths where flux does not peak (Figure 7 of \cite{Muller2014}). This strengthens the argument that simple TPM is sufficient to obtain the physical parameters using near infrared.

Another cause of uncertainty in the diameter is the change in cross-section due to the irregular shape. From \cite{Urakawa2011}, 4015WH has an elongated shape, i.e., 1.5:1.5:1.0 triaxial shape (non-tumbler, long axis mode, $ \beta_s = -27^{\circ} $; $ \beta_s $ slightly differ in tumbler model). Considering the near-zero orbital inclination of 4015WH ($ \sim 2.8^{\circ} $), we can calculate the cross-section difference to be $ \sim \pm 10 $--15 \%, which is directly proportional to the thermal flux.

\subsection{Sources of Uncertainties} \label{subsec: error sources}
In this subsection, we discuss the other sources of uncertainty. These are regarded as more fundamental sources of uncertainty because these sources arise from the ground base of the thermal model assumptions or from observational uncertainty.

The absolute magnitude may marginally affect the accuracy of the parameters obtained from TPM (see, e.g., section 2.8 of \citealt{Delbo2004} or section 6 of \citealt{Muller2005}). The absolute magnitude of HR30 was derived using $ G=0.0 $ and $ p_R =0.05 $ by \cite{Hicks2007}; thus, the uncertainty they provided (0.01 magnitude) should be regarded as a lower limit of the actual uncertainty. For 4015WH, uncertainty of $ H_V $ is 0.1 (Table \ref{tab: known parameters}), and it may increase if we consider the credibility of the standard $ H $, $ G $ system.

The rotational period of HR30 is also uncertain: 68 and 70.7 hours have also been proposed, as well as 73.2 hours \citep{Galad2008}. As mentioned in Section \ref{sec:Methodology}, a change in the rotational period will adjust the $ \Gamma $ value corresponding to the thermal parameter by a factor of $ \sqrt{P'/P} $ (0.964 and 0.983, respectively). Therefore, our estimation of the $ \Gamma $ value might be approximately 5 \% higher than the value obtained using a different rotational period. This is not a large uncertainty considering the previously published datasets (e.g., Fig 9 of \cite{Delbo2015a} or Fig \ref{fig:TIvsD}). 

HR30 possesses another interesting feature: None of the good-fitting models could reproduce the survey mode observations if we use the 2007-01-15 spectroscopic data. One possible scenario is that HR30 has irregular features in a certain region, and we observed different facets on 2007-01-15 (spectroscopy) and 2007-01-13 and 14 (survey mode). Considering the rotational period of HR30, which is nearly 3 days, this possibility is not rejected until more observations are made.

In our model, we neglected the seasonal effect. This may not be a good assumption, especially for objects with large rotational period. HR30 is such a case; the observations were made approximately 10 days ($ \sim 3 P_{\rm rot} $) after the perihelion (2007-01-02 06:50 UT), with true anomaly $ f \approx 10^{\circ} $. The true anomaly change rate was $ \dot{f} \approx 1^{\circ}/\text{day} $, and the heliocentric distance change rate was $ \dot{r}_h \approx 0.002 $ AU/day ($ r_h \approx 1.2 $ AU) when the observations were made. Therefore, the fractional change is about $ \dot{r_h}/r_h \approx 0.5$ \% per rotation. Considering $ P_{\rm rot} \approx 3 $ days, the basic assumption of TPM, i.e., thermal equilibrium, may not had been reached. As the HR30 observations were made \emph{after} the perihelion, the real $ \Gamma $ may be smaller than our fitted values because the thermal lag makes the night side appear hotter than thermal equilibrium \citep{Davidsson2009}. \cite{Davidsson2009} argued that Tempel 1 was observed \emph{before} the perihelion, and the seasonal effect was negligible since the model gave a higher temperature than the steady state. The slow change in its heliocentric distance also strengthens their argument: $ \dot{r}_h/r_h \approx 0.01 $ \% per rotation in early July 2005 when the Deep Impact observation was made, which is about 50 times smaller than that of HR30. Moreover, the rotational period of Tempel 1 is 40 hours, so the thermal equilibrium is reached $ \sim 2 $ times faster than that of HR30. Thus, it is not trivial to neglect seasonal effect for HR30.

The assumption of constant hemispherical emissivity over the thermal wavelengths ($ \varepsilon_h = 0.9 $) is also an issue since the model flux density is directly proportional to it. Another plausible assumption is $ \varepsilon_{h, {\rm Kir}} = 1- A_B $, based on Kirchhoff's law. For the slope parameter, $ G \in [0.0,~ 0.5]$ and the geometric albedo $ p_V \in [0.03,~ 0.07] $, we obtain $ \varepsilon_{h, {\rm Kir}} \approx 1 - A_B \sim 0.95-0.99 $ using Eq \ref{eq: Bond albedo}, which is $ \sim 5-10$ \% larger than the model value ($ \varepsilon_h = 0.9 $). This can be regarded as an error source for diameter determination.

\subsection{Physical Properties and Implications} \label{subsec: Implications}
As described in previous sections and can be seen in Fig \ref{fig: 4015 chisq} and \ref{fig:HR30_Pole_Projection}, low $ \chisq $ values are distributed throughout a certain domain in the parameter space, so it is difficult to pinpoint a single set of best-fit parameters. In the following paragraphs, we stress the importance of this work despite the limited dataset and model.

Firstly, the effective diameter, and thus the $ p_V $, is confined to very narrow ranges for both targets, despite the limited number of observational datasets. \cite{Muller2014} also showed that these parameters are well constrained, even when using a spherical model. For 4015WH, $ D = 3.7 $--4.4 km ($ p_V = 0.04 $--0.055) and $ \Gamma = 100 $--250 SI, with the minimum reduced chi-square statistic at $ D = 4.4 $ km with $ \Gamma = 250 $ SI.

From a taxonomic perspective, 4015WH is classified as a CF-type asteroid \citep{Tholen1989} or as a B-type asteroid following the Bus--DeMeo classification \citep{DeMeo2015}. \cite{DeMeo2013a} found that 833 B-type asteroids have an average $ p_V = 0.071 \pm 0.033 $ (Table 4 of their paper), which is consistent with our result ($ p_V = 0.04 $--0.055). Additionally, the size and albedo derived from our TPM is consistent with those of Spitzer Space Telescope infrared and ground based photometric observations: $ p_V = 0.059 \pm 0.011 $, $ D = 3.46 \pm 0.32 $ km from \cite{Licandro2009} and $ p_R = 0.055 \pm 0.012 $, $ D = 3.63 \pm 0.56$ km from \cite{Ishiguro2011}. The thermal inertia we obtained ($ \Gamma =$ 100--250 SI) is also consistent with the lower bound (60 SI) set by \cite{Licandro2009}. These facts indicate that the model assumptions in Section \ref{sec:Methodology} produce reliable results to some extent. We further decreased the uncertainty of both diameter and albedo compared to previous works. 

Although the physical quantities of HR30 are not known, it is impressive that the derived $ p_V $ value is strongly constrained to values consistent with those of typical cometary nuclei (0.02--0.06, \citealt{Lamy2004}). As for 4015WH, we emphasize that \emph{none} of the models with $ D > 27.1 $ km or $ D < 23.9 $ km ($ p_V < 0.035 $ or $ p_V > 0.045 $, respectively) gave good fitting (Fig \ref{fig:HR30_Pole_Projection}). Note that this value is robust against changes to other parameters, such as the pole and $ \Gamma $, even when a loose $ \chisq < 2 $ criterion is used. The $ \Gamma $ value is distributed over a wide range, 250--1,000 SI. Since the previously published value for the effective diameter $ D \sim 22 $ km was obtained by assuming $ p_R = 0.05 $ and $ G=0.0 $ \citep{Hicks2007}, our result is the first robust result for this target based on the high-quality AKARI IRC spectroscopic data. 

The pole orientation of HR30, which was not known a priori, could also be confined to a certain range: $ -20^{\circ} \lesssim \beta_s \lesssim 60 ^{\circ} $, i.e., near the ecliptic plane of the poles (Fig \ref{fig:HR30_Pole_Projection}). Furthermore, we can probabilistically conjecture that HR30 has $ \lambda_s \approx 45 ^{\circ} $ or $ \lambda_s \approx 135^{\circ} $, which differ by approximately $ 90^{\circ} $, though we cannot reject many other possible $ \lambda_s $ values. 

These derived values may be used as a priori knowledge in future TPM analysis of the target, thus reducing the volume in the parameter space and improving the efficiency of computing model fluxes. Although systematic studies on the quantitative reliability of pole orientation derived from a smooth spherical model have not been performed for a large number of small bodies, we conjecture that our pole solutions for HR30 may be used as constraints for future research on HR30, including studies using light curves. 

Finally, recent theoretical developments enable us to estimate the thermophysical parameters and surface particle size of asteroids using the $ \Gamma$ values. The thermal conductivity, $ \kappa $, can be estimated from Eq \ref{eq: TI def}. Assuming a bulk density $ \rho = 1400 $ kg/m$ ^3 $ (\citealt{Britt2002}; average for C-type asteroids) and $ c_s = 500 $ J/kg/K (\citealt{Opeil2010}; carbonaceous chondrites) for 4015WH, we obtain $ \kappa \sim 0.01-0.09 $ W/m/K and thermal skin depth $ l_s \sim 3-7$ cm (Eq \ref{eq: thermal skin depth}). Using the fitting function (Eq 9 of \citealt{Gundlach2013} with parameter set DS1 from their Table 6), we obtain the representative grain size on 4015WH of roughly 1--3 mm.

\section{Conclusion}
In this study, we applied the simple thermophysical model (TPM) described in Section \ref{sec:Methodology} to AKARI observations to investigate the physical properties of two comet-like targets: 107P/(4015) Wilson--Harrington (4015WH) and P/2006 HR30 (Siding Spring; HR30). The results can be summarized as follows.

\begin{enumerate}
	\item 4015WH, which is a potential future sample return mission target, was found to have effective diameter 3.74--4.39 km with geometric albedo 0.040--0.055, and thermal inertia 100--250 SI. The size and albedo are confined to narrower values than that of previous works (e.g., \citealt{Licandro2009,Ishiguro2011}) and is consistent with its spectral type (B- or CF-type). Under assumptions suitable for C-type asteroids, the surface grain size is estimated to be roughly 1--3 mm. 
	
	\item HR30, which is one of few known comets that have been spectrally observed with no detectable cometary activity, was found to have diameter 23.9--27.1 km and geometric albedo 0.035--0.045, which is consistent with many known cometary nuclei. The thermal inertia is estimated to be 250--1,000 SI with pole latitude $ - 20^{\circ} \lesssim \beta_s \lesssim +60^{\circ}$ and longitude most likely $ \lambda_s \approx 45^{\circ} $ or $ 135^{\circ} $. The possibility of irregular shape of the target is not rejected (Subsection \ref{subsec: error sources}). 
	
	\item Comet-like objects, although some possess slightly smaller $ \Gamma$ values than asteroidal counterparts, are not clearly distinguishable from normal asteroidal objects on the $ D $--$ \Gamma $ plane (Fig \ref{fig:TIvsD}).
\end{enumerate}

\acknowledgments
This research is based on observations from \emph{AKARI}, a JAXA project with the participation of ESA. This work was supported by the National Research Foundation of Korea (NRF) funded by the South Korean government (MEST; No. 2015R1D1A1A01060025).  This work was partly supported by JSPS KAKENHI Grant Number JP17K05636. The computation and plotting processes benefited greatly from NumPy \citep{Walt2011} and Matplotlib \citep{Hunter2007}. We appreciate Dr. Daisuke Ishihara at Nagoya University for providing the AKARI survey mode data of HR30.

\bibliography{library}

\begin{thebibliography}{}
\expandafter\ifx\csname natexlab\endcsname\relax\def\natexlab#1{#1}\fi
\providecommand{\url}[1]{\href{#1}{#1}}

\bibitem[{A'Hearn {et~al.}(2005)A'Hearn, Belton, Delamere, Kissel, Klaasen,
  McFadden, Meech, Melosh, Schultz, Sunshine, Thomas, Veverka, Yeomans, Baca,
  Busko, Crockett, Collins, Desnoyer, Eberhardy, Ernst, Farnham, Feaga,
  Groussin, Hampton, Ipatov, Li, Lindler, Lisse, Mastrodemos, Owen, Richardson,
  Wellnitz, \& White}]{AHearn2005}
A'Hearn, M.~F., Belton, M. J.~S., Delamere, W.~A., {et~al.} 2005, Science, 310,
  258

\bibitem[{Allen(1970)}]{Allen1970}
Allen, D.~A. 1970, Nature, 227, 158

\bibitem[{Baba {et~al.}(2016)Baba, Nakagawa, Shirahata, Isobe, Usui, Ohyama,
  Onaka, Yano, \& Kochi}]{Baba2016}
Baba, S., Nakagawa, T., Shirahata, M., {et~al.} 2016, Publ. Astron. Soc. Japan,
  68, 27

\bibitem[{Barucci {et~al.}(2009)Barucci, Yoshikawa, Michel, Kawagushi, Yano,
  Brucato, Franchi, Dotto, Fulchignoni, \& Ulamec}]{Barucci2009}
Barucci, M.~A., Yoshikawa, M., Michel, P., {et~al.} 2009, Exp. Astron., 23, 785

\bibitem[{Bottke {et~al.}(2002)Bottke, Morbidelli, Jedicke, Petit, Levison,
  Michel, \& Metcalfe}]{Bottke2002}
Bottke, W.~F., Morbidelli, A., Jedicke, R., {et~al.} 2002, Icarus, 156, 399

\bibitem[{Bowell {et~al.}(1989)Bowell, Hapke, Domingue, Lumme, Peltoniemi, \&
  Harris}]{Bowell1989}
Bowell, E., Hapke, B., Domingue, D., {et~al.} 1989, in Asteroids II, tucson
  edn., ed. R.~P. Binzel, T.~Gehrels, \& M.~S. Matthews (The University of
  Arizona Press), 525--556

\bibitem[{Britt {et~al.}(2002)Britt, Yeomans, Housen, \&
  Consolmagno}]{Britt2002}
Britt, D.~T., Yeomans, D., Housen, K., \& Consolmagno, G. 2002, in Asteroids
  III, ed. W.~F. Bottke, A.~Cellino, P.~Paolicchi, \& R.~P. Binzel (Tucson: The
  University of Arizona Press), 485--500

\bibitem[{Davidsson {et~al.}(2009)Davidsson, Guti{\'{e}}rrez, \&
  Rickman}]{Davidsson2009}
Davidsson, B. J.~R., Guti{\'{e}}rrez, P.~J., \& Rickman, H. 2009, Icarus, 201,
  335

\bibitem[{Davidsson {et~al.}(2013)Davidsson, Guti{\'{e}}rrez, Groussin,
  A'Hearn, Farnham, Feaga, Kelley, Klaasen, Merlin, Protopapa, Rickman,
  Sunshine, \& Thomas}]{Davidsson2013}
Davidsson, B. J.~R., Guti{\'{e}}rrez, P.~J., Groussin, O., {et~al.} 2013,
  Icarus, 224, 154

\bibitem[{Davidsson {et~al.}(2015)Davidsson, Rickman, Bandfield, Groussin,
  Guti{\'{e}}rrez, Wilska, Capria, Emery, Helbert, Jorda, Maturilli, \&
  Mueller}]{Davidsson2015}
Davidsson, B. J.~R., Rickman, H., Bandfield, J.~L., {et~al.} 2015, Icarus, 252,
  1

\bibitem[{Delbo(2004)}]{Delbo2004}
Delbo, M. 2004, PhD thesis, Free University of Berlin

\bibitem[{Delbo {et~al.}(2015)Delbo, Mueller, Emery, Rozitis, \&
  Capria}]{Delbo2015a}
Delbo, M., Mueller, M., Emery, J.~P., Rozitis, B., \& Capria, M.~T. 2015, in
  Asteroids IV, ed. P.~Michel, F.~E. DeMeo, \& W.~F. Bottke (Tucson: The
  University of Arizona Press), 107--128

\bibitem[{Delbo {et~al.}(2014)Delbo, Libourel, Wilkerson, Murdoch, Michel,
  Ramesh, Ganino, Verati, \& Marchi}]{Delbo2014}
Delbo, M., Libourel, G., Wilkerson, J., {et~al.} 2014, Nature, 508, 233

\bibitem[{DeMeo {et~al.}(2015)DeMeo, Alexander, {Walsh K. J.}, Chapman, \&
  Binzel}]{DeMeo2015}
DeMeo, F.~E., Alexander, C. M.~O., {Walsh K. J.}, Chapman, C.~R., \& Binzel,
  R.~P. 2015, in Asteroids IV, ed. P.~Michel, F.~E. DeMeo, \& W.~F. Bottke
  (Tucson: The University of Arizona Press), 13--41

\bibitem[{DeMeo \& Carry(2013)}]{DeMeo2013a}
DeMeo, F.~E., \& Carry, B. 2013, Icarus, 226, 723

\bibitem[{Dickel(1979)}]{Dickel1979}
Dickel, J.~R. 1979, in Asteroids, ed. T.~Gehrels (Tucson: The University of
  Arizona Press), 212--221

\bibitem[{Fern{\'{a}}ndez {et~al.}(1997)Fern{\'{a}}ndez, McFadden, Lisse,
  Helin, \& Chamberlin}]{Fernandez1997}
Fern{\'{a}}ndez, Y.~R., McFadden, L.~A., Lisse, C.~M., Helin, E.~F., \&
  Chamberlin, A.~B. 1997, Icarus, 128, 114

\bibitem[{Fowler \& Chillemi(1992)}]{Fowler1992}
Fowler, J.~W., \& Chillemi, J.~R. 1992, in IRAS Minor Planet Surv., ed. E.~F.
  Tedesco, G.~J. Veeder, J.~W. Fowler, \& J.~R. Chillemi (Phillips Laboratory
  Technical Report No.PL-TR-92-2049), 17--43

\bibitem[{Galad(2008)}]{Galad2008}
Galad, A. 2008, Minor Planet Bull. (ISSN 1052-8091). Bull. Minor Planets Sect.
  Assoc. Lunar Planet. Obs. Vol. 35, No. 1, p. 17-21, 35, 17

\bibitem[{Groussin {et~al.}(2013)Groussin, Sunshine, Feaga, Jorda, Thomas, Li,
  A'Hearn, Belton, Besse, Carcich, Farnham, Hampton, Klaasen, Lisse, Merlin, \&
  Protopapa}]{Groussin2013}
Groussin, O., Sunshine, J., Feaga, L., {et~al.} 2013, Icarus, 222, 580

\bibitem[{Gulkis {et~al.}(2015)Gulkis, Allen, von Allmen, Beaudin, Biver,
  Bockel{\'{e}}e-Morvan, Choukroun, Crovisier, Davidsson, Encrenaz, Encrenaz,
  Frerking, Hartogh, Hofstadter, Ip, Janssen, Jarchow, Keihm, Lee, Lellouch,
  Leyrat, Rezac, Schloerb, \& Spilker}]{Gulkis2015a}
Gulkis, S., Allen, M., von Allmen, P., {et~al.} 2015, Science, 347, aaa0709

\bibitem[{Gundlach \& Blum(2013)}]{Gundlach2013}
Gundlach, B., \& Blum, J. 2013, Icarus, 223, 479

\bibitem[{Hanu{\v{s}} {et~al.}(2015)Hanu{\v{s}}, Delbo', \v{D}urech,
  Al{\'{i}}-Lagoa, Durech, \& Al{\'{i}}-Lagoa}]{Hanus2015}
Hanu{\v{s}}, J., Delbo', M., \v{D}urech, J., {et~al.} 2015, Icarus, 256, 101

\bibitem[{Harada(2009)}]{Harada2009}
Harada, A. 2009, PhD thesis, The University of Tokyo

\bibitem[{Harris(1998)}]{Harris1998}
Harris, A.~W. 1998, Icarus, 131, 291

\bibitem[{Hicks \& Bauer(2007)}]{Hicks2007}
Hicks, M.~D., \& Bauer, J.~M. 2007, Astrophys. J., 662, L47

\bibitem[{Hunter(2007)}]{Hunter2007}
Hunter, J.~D. 2007, Comput. Sci. Eng., 9, 90

\bibitem[{Ishiguro {et~al.}(2011)Ishiguro, Ham, Tholen, Elliott, Micheli, Niwa,
  Sakamoto, Matsuda, Urakawa, Yoshimoto, Sarugaku, Usui, Hasegawa, Iwata,
  Ozaki, Kuroda, \& Ootsubo}]{Ishiguro2011}
Ishiguro, M., Ham, J.-B., Tholen, D.~J., {et~al.} 2011, Astrophys. J., 726, 101

\bibitem[{Kaasalainen(2011)}]{Kaasalainen2011}
Kaasalainen, M. 2011, Inverse Probl. Imaging, 5, 37

\bibitem[{Kaplinger {et~al.}(2013)Kaplinger, Wie, \& Dearborn}]{Kaplinger2013}
Kaplinger, B., Wie, B., \& Dearborn, D. 2013, Acta Astronaut., 90, 156

\bibitem[{Kawaguchi(2002)}]{Kawaguchi2002}
Kawaguchi, J. 2002, Adv. Sp. Res., 29, 1215

\bibitem[{Kim {et~al.}(2014)Kim, Ishiguro, \& Usui}]{Kim2014}
Kim, Y., Ishiguro, M., \& Usui, F. 2014, Astrophys. J., 789, 151

\bibitem[{Lamy {et~al.}(2004)Lamy, Toth, Fernandez, \& Weaver}]{Lamy2004}
Lamy, P.~L., Toth, I., Fernandez, Y.~R., \& Weaver, H.~A. 2004, in Comets II,
  ed. {M. C. Festou}, {H. U. Keller}, \& {H. A. Weaver} (Tucson: University of
  Arizona Press), 223--264

\bibitem[{Lebofsky \& Spencer(1989)}]{Lebofsky1989}
Lebofsky, L.~A., \& Spencer, J.~R. 1989, in Asteroids II, ed. R.~P. Binzel,
  T.~Gehrels, \& M.~S. Matthews (Tucson: The University of Arizona Press),
  128--147

\bibitem[{Lebofsky {et~al.}(1986)Lebofsky, Sykes, Tedesco, Veeder, Matson,
  Brown, Gradie, Feierberg, \& Rudy}]{Lebofsky1986a}
Lebofsky, L.~A., Sykes, M.~V., Tedesco, E.~F., {et~al.} 1986, Icarus, 68, 239

\bibitem[{Licandro {et~al.}(2009)Licandro, Campins, Kelley, Fern{\'{a}}ndez,
  Delb{\'{o}}, Reach, Groussin, Lamy, Toth, A'Hearn, Bauer, Lowry, Fitzsimmons,
  Lisse, Meech, Pittichov{\'{a}}, Snodgrass, \& Weaver}]{Licandro2009}
Licandro, J., Campins, H., Kelley, M., {et~al.} 2009, Astron. Astrophys., 507,
  1667

\bibitem[{Mainzer {et~al.}(2011)Mainzer, Bauer, Grav, Masiero, Cutri, Dailey,
  Eisenhardt, McMillan, Wright, Walker, Jedicke, Spahr, Tholen, Alles, Beck,
  Brandenburg, Conrow, Evans, Fowler, Jarrett, Marsh, Masci, McCallon,
  Wheelock, Wittman, Wyatt, DeBaun, Elliott, Elsbury, Gautier, Gomillion,
  Leisawitz, Maleszewski, Micheli, \& Wilkins}]{Mainzer2011c}
Mainzer, A., Bauer, J.~M., Grav, T., {et~al.} 2011, Astrophys. J., 731, 53

\bibitem[{Morrison \& Lebofsky(1979)}]{Morrison1979}
Morrison, D., \& Lebofsky, L. 1979, in Asteroids, ed. T.~Gehrels (Tucson: The
  University of Arizona Press), 184--205

\bibitem[{Mueller(2007)}]{Mueller2007}
Mueller, M. 2007, PhD thesis, Free University of Berlin, arXiv:1208.3993

\bibitem[{Mueller {et~al.}(2010)Mueller, Marchis, Emery, Harris, Mottola,
  Hestroffer, Berthier, \& di~Martino}]{Mueller2010}
Mueller, M., Marchis, F., Emery, J.~P., {et~al.} 2010, Icarus, 205, 505

\bibitem[{M{\"{u}}ller {et~al.}(2014)M{\"{u}}ller, Hasegawa, \&
  Usui}]{Muller2014}
M{\"{u}}ller, T.~G., Hasegawa, S., \& Usui, F. 2014, Publ. Astron. Soc. Japan,
  66, 52

\bibitem[{M{\"{u}}ller {et~al.}(2005)M{\"{u}}ller, Sekiguchi, Kaasalainen, Abe,
  \& Hasegawa}]{Muller2005}
M{\"{u}}ller, T.~G., Sekiguchi, T., Kaasalainen, M., Abe, M., \& Hasegawa, S.
  2005, Astron. Astrophys., 443, 347

\bibitem[{M{\"{u}}ller {et~al.}(2017)M{\"{u}}ller, Ďurech, Ishiguro, Mueller,
  Kr{\"{u}}hler, Yang, Kim, O'Rourke, Usui, Kiss, Altieri, Carry, Choi, Delbo,
  Emery, Greiner, Hasegawa, Hora, Knust, Kuroda, Osip, Rau, Rivkin, Schady,
  Thomas-Osip, Trilling, Urakawa, Vilenius, Weissman, \& Zeidler}]{Muller2017}
M{\"{u}}ller, T.~G., Ďurech, J., Ishiguro, M., {et~al.} 2017, Astron.
  Astrophys., 599, A103

\bibitem[{Murakami {et~al.}(2007)Murakami, Baba, Barthel, Clements, Cohen, Doi,
  Enya, Figueredo, Fujishiro, Fujiwara, Fujiwara, Garcia-Lario, Goto, Hasegawa,
  Hibi, Hirao, Hiromoto, Hong, Imai, Ishigaki, Ishiguro, Ishihara, Ita, Jeong,
  Jeong, Kaneda, Kataza, Kawada, Kawai, Kawamura, Kessler, Kester, Kii, Kim,
  Kim, Kobayashi, Koo, Kwon, Lee, Lorente, Makiuti, Matsuhara, Matsumoto,
  Matsuo, Matsuura, M{\"{u}}ller, Murakami, Nagata, Nakagawa, Naoi, Narita,
  Noda, Oh, Ohnishi, Ohyama, Okada, Okuda, Oliver, Onaka, Ootsubo, Oyabu, Pak,
  Park, Pearson, Rowan-Robinson, Saito, Sakon, Salama, Sato, Savage, Serjeant,
  Shibai, Shirahata, Sohn, Suzuki, Takagi, Takahashi, Tanabe, Takeuchi, Takita,
  Thomson, Uemizu, Ueno, Usui, Verdugo, Wada, Wang, Watabe, Watarai, White,
  Yamamura, Yamauchi, \& Yasuda}]{Murakami2007}
Murakami, H., Baba, H., Barthel, P., {et~al.} 2007, Publ. Astron. Soc. Japan,
  59, S369

\bibitem[{Myhrvold(2016)}]{Myhrvold2016a}
Myhrvold, N. 2016, Publ. Astron. Soc. Pacific, 128, 045004

\bibitem[{Ohyama {et~al.}(2007)Ohyama, Onaka, Matsuhara, Wada, Kim, Fujishiro,
  Uemizu, Sakon, Cohen, Ishigaki, Ishihara, Ita, Kataza, Matsumoto, Murakami,
  Oyabu, Tanab{\'{e}}, Takagi, Ueno, Usui, Watarai, Pearson, Takeyama,
  Yamamuro, \& Ikeda}]{Ohyama2007}
Ohyama, Y., Onaka, T., Matsuhara, H., {et~al.} 2007, Publ. Astron. Soc. Japan,
  59, S411

\bibitem[{Onaka {et~al.}(2007)Onaka, Matsuhara, Wada, Fujishiro, Fujiwara,
  Ishigaki, Ishihara, Ita, Kataza, Kim, Matsumoto, Murakami, Ohyama, Oyabu,
  Sakon, Tanab{\'{e}}, Takagi, Uemizu, Ueno, Usui, Watarai, Cohen, Enya,
  Ootsubo, Pearson, Takeyama, Yamamuro, \& Ikeda}]{Onaka2007}
Onaka, T., Matsuhara, H., Wada, T., {et~al.} 2007, Publ. Astron. Soc. Japan,
  59, S401

\bibitem[{Onaka {et~al.}(2010)Onaka, Matsuhara, Wada, Ishihara, Ita, Ohyama,
  Ootsubo, Oyabu, Sakon, Shimonishi, Takita, Tanab{\`{e}}, Usui, \&
  Murakami}]{Onaka2010}
Onaka, T., Matsuhara, H., Wada, T., {et~al.} 2010, in Sp. Telesc. Instrum. 2010
  Opt. Infrared, Millim. Wave., ed. J.~M. {Oschmann, Jr.}, M.~C. Clampin, \&
  H.~A. MacEwen, Vol. 7731 (Proceedings of the SPIE), 77310M

\bibitem[{Ootsubo {et~al.}(2012)Ootsubo, Kawakita, Hamada, Kobayashi,
  Yamaguchi, Usui, Nakagawa, Ueno, Ishiguro, Sekiguchi, Watanabe, Sakon,
  Shimonishi, \& Onaka}]{Ootsubo2012a}
Ootsubo, T., Kawakita, H., Hamada, S., {et~al.} 2012, Astrophys. J., 752, 15

\bibitem[{Opeil {et~al.}(2010)Opeil, Consolmagno, \& Britt}]{Opeil2010}
Opeil, C., Consolmagno, G., \& Britt, D. 2010, Icarus, 208, 449

\bibitem[{Planck(1914)}]{Planck1914}
Planck, M. 1914, {The Theory of Heat Radiation}, 2nd edn. (Philadelphia: P.
  Blakiston's Son {\&} Co.), doi:10.1038/123755a0

\bibitem[{Pravec \& Harris(2007)}]{Pravec2007}
Pravec, P., \& Harris, A.~W. 2007, Icarus, 190, 250

\bibitem[{Press {et~al.}(2007)Press, Teukolsky, Vetterling, \&
  Flannery}]{Press2007}
Press, W.~H., Teukolsky, S.~A., Vetterling, W.~T., \& Flannery, B.~P. 2007,
  {Numerical Recipes The Art of Scientific Computing}, 3rd edn. (Cambridge
  University Press)

\bibitem[{Putzig(2006)}]{Putzig2006}
Putzig, N.~E. 2006, PhD thesis, University of Colorado

\bibitem[{Schorghofer(2008)}]{Schorghofer2008}
Schorghofer, N. 2008, Astrophys. J., 682, 697

\bibitem[{Shi {et~al.}(2016)Shi, Hu, Sierks, G{\"{u}}ttler, A'Hearn, Blum,
  El-Maarry, K{\"{u}}hrt, Mottola, Pajola, Oklay, Fornasier, Tubiana, Keller,
  Vincent, Bodewits, H{\"{o}}fner, Lin, Gicquel, Hofmann, Barbieri, Lamy,
  Rodrigo, Koschny, Rickman, Barucci, Bertaux, Bertini, Cremonese, {Da Deppo},
  Davidsson, Debei, {De Cecco}, Fulle, Groussin, Gutierrez, Hviid, Ip, Jorda,
  Knollenberg, Kovacs, Kramm, K{\"{u}}ppers, Lara, Lazzarin, {Lopez Moreno},
  Marzari, Naletto, \& Thomas}]{Shi2016}
Shi, X., Hu, X., Sierks, H., {et~al.} 2016, Astron. Astrophys., 586, A7

\bibitem[{Shimonishi {et~al.}(2013)Shimonishi, Onaka, Kato, Sakon, Ita,
  Kawamura, \& Kaneda}]{Shimonishi2013}
Shimonishi, T., Onaka, T., Kato, D., {et~al.} 2013, Astron. J., 145, 32

\bibitem[{Spencer {et~al.}(1989)Spencer, Lebofsky, \& Sykes}]{Spencer1989}
Spencer, J.~R., Lebofsky, L.~A., \& Sykes, M.~V. 1989, Icarus, 78, 337

\bibitem[{Tedesco {et~al.}(2002)Tedesco, Noah, Noah, \& Price}]{Tedesco2002}
Tedesco, E.~F., Noah, P.~V., Noah, M., \& Price, S.~D. 2002, Astron. J., 123,
  1056

\bibitem[{Tholen(1989)}]{Tholen1989}
Tholen, D.~J. 1989, in Asteroids II, ed. R.~P. Binzel, T.~Gehrels, \& M.~S.
  Matthews (Tucson: The University of Arizona Press), 1139--1150

\bibitem[{Urakawa {et~al.}(2011)Urakawa, Okumura, Nishiyama, Sakamoto,
  Takahashi, Abe, Ishiguro, Kitazato, Kuroda, Hasegawa, Ohta, Kawai, Shimizu,
  Nagayama, Yanagisawa, Yoshida, \& Yoshikawa}]{Urakawa2011}
Urakawa, S., Okumura, S.-i., Nishiyama, K., {et~al.} 2011, Icarus, 215, 17

\bibitem[{Usui {et~al.}(2011)Usui, Kuroda, M{\"{u}}ller, Hasegawa, Ishiguro,
  Ootsubo, Ishihara, Kataza, Takita, Oyabu, Ueno, Matsuhara, \&
  Onaka}]{Usui2011a}
Usui, F., Kuroda, D., M{\"{u}}ller, T.~G., {et~al.} 2011, Publ. Astron. Soc.
  Japan, 63, 1117

\bibitem[{van~der Walt {et~al.}(2011)van~der Walt, Colbert, \&
  Varoquaux}]{Walt2011}
van~der Walt, S., Colbert, S.~C., \& Varoquaux, G. 2011, Comput. Sci. Eng., 13,
  22

\bibitem[{Vokrouhlicky {et~al.}(2015)Vokrouhlicky, Bottke, Chesley, Scheeres,
  \& Statler}]{Vokrouhlicky2015}
Vokrouhlicky, D., Bottke, W.~F., Chesley, S.~R., Scheeres, D.~J., \& Statler,
  T.~S. 2015, in Asteroids IV, ed. P.~Michel, F.~E. DeMeo, \& W.~F. Bottke
  (Tucson: The University of Arizona Press), 509--531

\bibitem[{Wie(2013)}]{Wie2013}
Wie, B. 2013, Acta Astronaut., 90, 146

\bibitem[{Yoshikawa {et~al.}(2008)Yoshikawa, Yano, \&
  Kawaguchi}]{Yoshikawa2008}
Yoshikawa, M., Yano, H., \& Kawaguchi, J. 2008, 39th Lunar Planet. Sci. Conf.,
  1747

\end{thebibliography}

\end{document}